\documentclass[a4paper,fleqn,usenatbib]{mnras}

\usepackage[T1]{fontenc}
\usepackage{ae,aecompl}
\usepackage{graphicx}
\usepackage{adjustbox}
\usepackage{pgfplots}
\usepackage{mathtools}
\usepackage{epstopdf}
\usepackage{hyperref}
\usepackage{graphicx}	
\usepackage{amsmath,bm}	
\usepackage{amssymb}	
\usepackage{breqn}
\usepackage{xcolor}

\makeatletter
\newcommand{\labitem}[2]{%
\def\@itemlabel{\textbf{#1}}
\item
\def\@currentlabel{#1}\label{#2}}
\makeatother

\title[The long-term scintillation of PSR J1141$-$6545]{Modelling annual and orbital variations in the scintillation of the relativistic binary PSR J1141$-$6545}

\author[D. J. Reardon et al.]{D. J. Reardon,$^{1,2,3}$\thanks{E-mail: dreardon@swin.edu.au}
W. A. Coles,$^{4}$
G. Hobbs,$^{3}$
S. Ord,$^{3}$
M. Kerr,$^{5}$
M. Bailes,$^{1}$\newauthor  
N. D. R. Bhat,$^{6}$
V. Venkatraman Krishnan$^{1,7}$
\\
$^{1}$Centre for Astrophysics and Supercomputing, Swinburne University of Technology, P.O. Box 218, Hawthorn, Victoria 3122,\\ Australia\\
$^{2}$Monash Centre for Astrophysics (MoCA), School of Physics and Astronomy, Monash University, Victoria 3800, Australia\\
$^{3}$Australia Telescope National Facility, CSIRO Astronomy \& Space Science, P.O. Box 76, Epping, NSW 1710, Australia\\
$^{4}$Electrical and Computer Engineering, University of California at San Diego, La Jolla, California, U.S.A.\\
$^{5}$Space Science Division, Naval Research Laboratory, Washington, DC 20375-5352, USA\\
$^{6}$International Centre for Radio Astronomy Research, Curtin University, Bentley, Western Australia 6102, Australia\\
$^{7}$Max-Planck-Institut fu\"r Radioastronomie, Auf dem Hu\"gel 69, 53121 Bonn, Germany \\
}

\date{\today}

\pubyear{2018}

\begin{document}
\label{firstpage}
\pagerange{\pageref{firstpage}--\pageref{lastpage}}
\maketitle

\begin{abstract}

We have observed the relativistic binary pulsar PSR J1141$-$6545 over a period of $\sim$6\,years using the Parkes 64\,m radio telescope, with a focus on modelling the diffractive intensity scintillations to improve the accuracy of the astrometric timing model. The long-term scintillation, which shows orbital and annual variations, allows us to measure parameters that are difficult to measure with pulsar timing alone. These include: the orbital inclination $i$; the longitude of the ascending node $\Omega$; and the pulsar system transverse velocity. We use the annual variations to resolve the previous ambiguity in the sense of the inclination angle. Using the correct sense, and a prior probability distribution given by a constraint from pulsar timing ($i=73\pm3^\circ$), we find $\Omega=24.8\pm1.8^\circ$ and we estimate the pulsar distance to be $D=10^{+4}_{-3}$\,kpc. This then gives us an estimate of this pulsar's proper motion of $\mu_{\alpha}\cos{\delta}=2.9\pm1.0$\,mas\,yr$^{-1}$ in right ascension and $\mu_{\delta}=1.8\pm0.6$\,mas\,yr$^{-1}$ in declination. Finally, we obtain measurements of the spatial structure of the interstellar electron density fluctuations, including: the spatial scale and anisotropy of the diffraction pattern; the distribution of scattering material along the line of sight; and spatial variation in the strength of turbulence from epoch to epoch. We find that the scattering is dominated by a thin screen at a distance of $(0.724\pm0.008)D$, with an anisotropy axial ratio $A_{\rm r} = 2.14\pm0.11$.
\end{abstract}

\begin{keywords}
pulsars: general, pulsars: individual (PSR J1141$-$6545), ISM: general, ISM: structure
\end{keywords}




\section{Introduction}
PSR J1141$-$6545 is a relativistic binary pulsar in a $\sim$4.7\,hour eccentric orbit with a young, white dwarf companion \citep[discovered by][]{Kaspi+00}. This system has proved to be a unique and valuable laboratory for testing general relativity (GR) in such asymmetrical mass systems \citep{Bhat+08, Manchester+10}, and its runaway velocity is interesting for investigating their formation and binary evolution mechanisms \citep[e.g.][]{Tauris+00, Davies+02, Church+06}. The short orbital period and large centripetal acceleration of this pulsar makes it ideal for measuring and modelling changes in its scintillation timescale \citep{Ord+02a} for the purpose of measuring astrometric parameters. Parameters including the inclination angle $i$ (and its sense), the longitude of ascending node $\Omega$, and the transverse velocity are useful for tests of GR and formation scenarios. We are able to uniquely determine these parameters and estimate the pulsar distance and proper motion by modelling the long-term scintillation of PSR J1141$-$6545 and accounting for anisotropy in the scattering. This is only possible by exploiting the annual variations observed with long-term monitoring of the pulsar.

Intensity variations caused by interstellar scintillation are seen in all observations of radio pulsars at centimetre to metre wavelengths. They are caused by transverse fluctuations in the electron density of the turbulent ionised interstellar medium (IISM). The primary mechanism is interference by waves scattered through diffraction \citep{Rickett69}. The diffractive scintillations are modulated by refractive scintillations on larger spatial scales \citep{Rickett+84, Romani+86}. 

The observed scintillations, which appear as variations in the source flux with time and observing frequency, are caused by the spatial diffraction pattern drifting across the line-of-sight. Thus the time scale of the observed scintillations $\tau_{\rm d}$ (typically of order minutes) is inversely proportional to this drift velocity $V_{\rm los}$, which is the velocity at which the line-of-sight crosses the IISM (Section \ref{sec:model}). The diffraction pattern is  frequency-dependent and becomes decorrelated over a bandwidth $\Delta\nu_{\rm d}$. The angular scattering broadens each pulse into a quasi-exponential pulse with timescale $\tau_{\rm s}$, which is related to the bandwidth by $2\pi\Delta\nu_{\rm d}\tau_{\rm s}\approx1$ \citep{Rickett77}. Strong diffractive scintillations have a narrow bandwidth (typically of order MHz), which is a useful measure of the strength of scattering and can be used to estimate the diffractive spatial scale $s_{\rm d}$ \citep{Cordes+98}. Detailed overviews of pulsar scintillations are given by \citet{Rickett90} and \citet{Narayan92}.

Measurements of a dynamic spectrum of intensity scintillation with time and frequency can therefore provide information on the spatial structure of the IISM, the transverse velocity of the pulsar, and the strength of scattering. Although scattering occurs throughout the line-of-sight, it is often dominated by one, or a few, local regions of more intense scattering.

For solitary pulsars $V_{\rm los}$ is often dominated by the pulsar proper motion and is relatively simple to model, depending only on the distance from the scattering region to the Earth \citep{Lyne+82}. If the pulsar has a binary companion, orbital dynamics can also be studied from the transverse orbital motion. This was first used by \citet{Lyne84} to measure the orbital inclination angle of PSR B0655$+$64 for the first time. Then \citet{Ord+02a} analysed two consecutive orbits of PSR J1141$-$6545 and found that the time scale varied smoothly over the orbit, but since they had to assume that the scattering was isotropic the parameter estimates likely suffered a bias. The technique was extended to deal with anisotropic scattering to analyse scintillations of the double pulsar PSR J0737$-$3039A by \citet{Coles+05}. It was further extended by \citet{Rickett+14} to analyze multiple observations of PSR J0737$-$3039A over several years, which requires including the variation of the Earth's velocity into the model. We have been provided access to the latter data and used it to calibrate our analysis against that of \citet{Rickett+14}. 

For this work we use the dynamic spectrum as our basic observable and model the diffractive scintillations of PSR J1141$-$6545. The physical parameters measured in this way are important for testing a variety of theories such as GR and other boost-invariant theories of gravity \citep{Damour+92}. However, poor timing precision currently limits the measurement of the Shapiro delay, which would otherwise provide a measurement of the inclination angle and companion mass to further constrain the tests of GR \citep{Bhat+08}. An independent measurement of the inclination angle from scintillation will provide complementary tests of GR. In addition, the distance to the pulsar is poorly constrained at present, so the contamination from kinematic effects \citep[in this case the Shklovskii effect; ][]{Shklovskii70} in the measured relativistic orbital parameters is unknown. A lower-bound distance estimate of 3.7\,kpc was found by \citet{Ord+02b} using \textsc{HI} absorption spectra and this was used in the scintillation modelling \citet{Ord+02a}. However \citet{Verbiest+12} showed that distances derived in this way may be overestimated because of a luminosity bias, and presented a revised distance estimate of $3\pm2$\,kpc (we take this as an initial value for our models). At this large distance, the proper motion is too small to be measured with current timing precision. Further, the white dwarf companion was not identified in a recent targeted search for binary pulsar companions \citep{Jennings+18} in the second \textit{Gaia} data release \citep{Gaia18}, which could have provided independent estimates of distance and proper motion. For these reasons, both the measured transverse velocity (to estimate the Shklovskii effect) and inclination angle (to constrain the pulsar and companion masses) from scintillation modelling are particularly useful for improving the tests of GR with this system.
 
In this paper we present new short-term and long-term scintillation models for different scattering geometries, which are described in Section \ref{sec:model}. We compare the evidence for each of these geometries using Bayesian methods in Section \ref{sec:fitandcompare}. Using the short-term models, in Section \ref{sec:singleepochs} we demonstrate that the Earth's velocity is detectable from scintillation timescale modulation over a year, and that the relativistic advance of periastron can be measured from scintillation alone. Our best long-term model, presented in Section \ref{sec:anisotropic} is a thin scattering screen and includes anisotropy in the IISM. The annual variation in this model allows us to measure the orientation of the pulsar's orbit in celestial coordinates, resolve the sense of the inclination angle, constrain the scattering anisotropy, determine the distance to the scattering region, and estimate the proper motion of the pulsar. We compare these results with previous measurements from scintillation and pulsar timing, provide a revised distance estimate to the pulsar in Section \ref{sec:distance}, and predict the contamination in the orbital period-derivative from the Shklovskii effect in Section \ref{sec:timing}.

\section{Dataset}
\label{sec:dataset}
\subsection{Observations and dynamic spectra}
\label{sec:obs}

\begin{figure*}
\includegraphics[trim={1cm 0.5cm 2.5cm 3cm},clip,width=\textwidth]{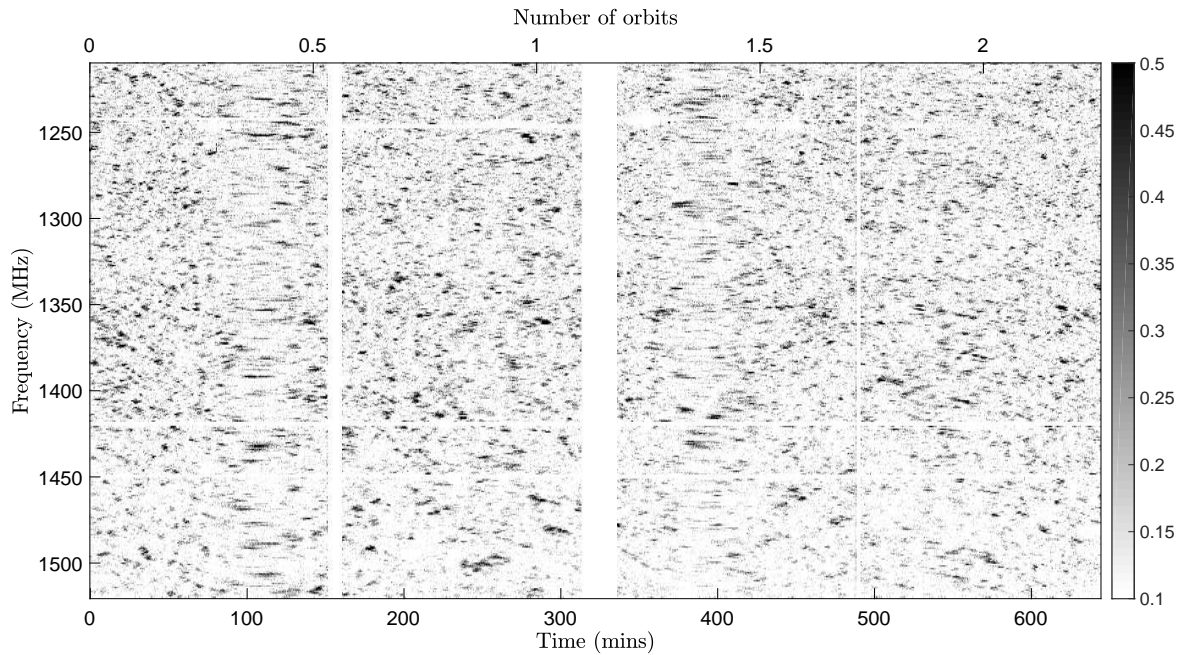}
\caption{A dynamic spectrum of four PSR J1141$-$6545 observations made on MJD 56391. The four observations cover a combined $\sim2.2$ pulsar orbits. Stretching of the scintles (in black) in time is evident (at e.g. 100 and 400\,minutes) when the pulsar orbital velocity reaches a minimum. The vertical white bars are periods between the individual observations, while horizontal white bars and patches were removed because of radio-frequency interference. The greyscale shows the normalised flux with the black and white limits chosen to optimise the visualisation of scintles.}
\label{fig:dynspec}
\end{figure*}

For this work, we use a selection of archival PSR J1141$-$6545 observations from the Parkes 64\,m radio telescope, spanning $\sim$6\,years from June 2009 to June 2015. The observations were part of the P361 observing project, which was a long-term campaign to improve the stringent tests of gravitational theories. The data were received with the central beam of the Parkes 20 cm multibeam receiver, and recorded either with a digital polyphase filterbank system with a 256\,MHz bandwidth and 0.25\,MHz channel width, or a coherent dedispersion machine with a 400\,MHz bandwidth and 0.78\,MHz channel width. We selected only the observations which spanned at least 142\,minutes, so that each observation covered at least half of an orbit. This yielded 126 individual observations which we treated as 23 distinct ``epochs" separated by at least 60 days. 

A dynamic spectrum for each observation was produced using the data processing pipeline designed for the upcoming data release 2 (DR2) of the Parkes Pulsar Timing Array \citep[PPTA;][]{Manchester+13} project. In brief, observations of a pulsed noise diode that excites both X and Y polarisations in phase are performed before each of the observations described above, to allow correction of the complex gain.  The noise diode is itself calibrated to absolute flux density using on- and off-source observations of the bright radio galaxy Hydra A. Polarisation calibration is done using the noise diode observations combined with regular observations of the highly polarised pulsar PSR J0437$-$4715. To compute dynamic spectra, we perform a least-squares fit of an analytic model of the pulse profile in total intensity (Stokes I) to the observed pulse profile for each sub-integration and frequency channel. This fit provides the amplitude and its uncertainty.  Because the observed pulsar profiles are already absolutely calibrated, the amplitude measurement yields the pulsar flux density directly. In addition, using this analytic pulse profile simultaneously optimises the signal-to-noise ratio for both the pulse amplitude and time of arrival. The calibrations and measurements described above are performed with the \textsc{psrchive} \citep{Hotan+04} package. 

Figure \ref{fig:dynspec} shows the dynamic spectra of four consecutive observations of PSR J1141$-$6545 on MJD 56391. The modulation of the scintillation timescale $\tau_{\rm d}$ caused by the orbital motion of the pulsar is visible by eye. We characterise the statistics of this spectrum by a two dimensional autocovariance function (ACF). Following convention \citep{Cordes+98}, the half-width at half maximum of this ACF in frequency is the decorrelation bandwidth $\Delta\nu_{\rm d}$ and the half-width at $1/e$ in time is $\tau_{\rm d}$. The scale of spatial variations in the diffraction pattern $s_{\rm d}$ is related to the scintillation timescale by $\tau_{\rm d} = s_{\rm d}/V_{\rm los}$. The spatial scale is determined by the strength of scattering, which can vary with time but can be estimated from $\Delta\nu_{\rm d}$. If the spatial structure is anisotropic, $\tau_{\rm d}$ will depend on the angle between the semi-major axis of the anisotropy and the velocity vector, which our long-term data is sensitive to. 

Fortunately, since $\Delta\nu_{\rm d}$ is a direct measure of the strength of scattering, this can be used to correct for temporal variations in the spatial scale caused by changes in the strength of scattering, provided that the anisotropy does not also change with time. A formalism for using $\Delta\nu_{\rm d}$ to correct for changes in $s_{\rm d}$ has been provided by \citet{Cordes+98}. The near-constant $\Delta\nu_{\rm d}$ for the dynamic spectrum in Figure \ref{fig:dynspec} shows that the strength of scattering does not change significantly over the orbital period. This is because the projected size of the orbit is smaller than the scattering disk. However we do observe small strength of scattering changes from epoch to epoch (discussed in Section \ref{sec:inhomogeneity}) and we use the scheme of \citet{Cordes+98} to account for this.

\begin{figure*}
\includegraphics[width=\textwidth]{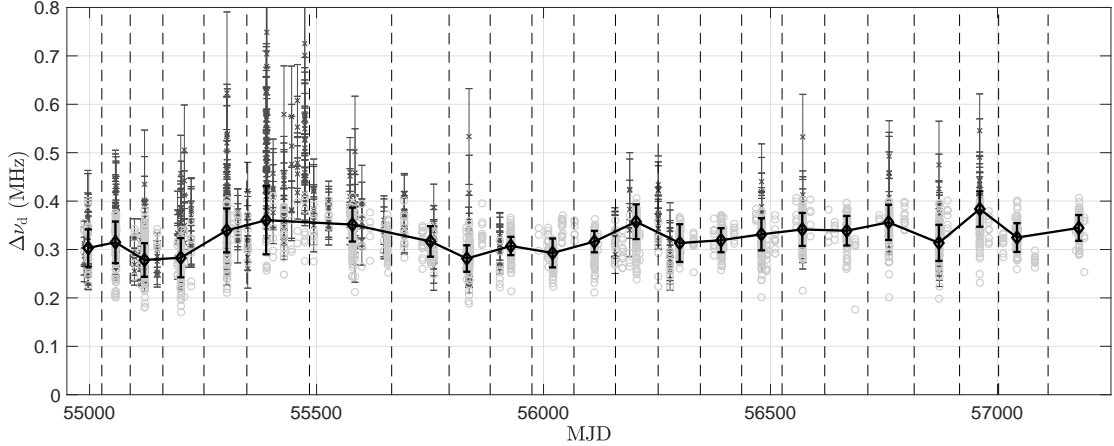}
\caption{Decorrelation bandwidth ($\Delta\nu_{\rm d}$) as a function of time for all archival observations in a $\sim$6\,year span with a total observing time of at least 24\,minutes at $\sim$1400\,MHz for PSR J1141$-$6545. The data are split into 23 epochs separated by vertical, dashed lines. Each epoch contains multiple individual observations, and the dynamic spectra for these were cut into $\sim$11\,minute segments. Black crosses are the measured $\Delta\nu_{\rm d}$ for each segment where $B_{\rm c} < \Delta\nu_{\rm d}$ for channel bandwidth $B_{\rm c}$.  Grey circles are a flux-based estimate of $\Delta\nu_{\rm d}$ from the procedure described in Section \ref{sec:inhomogeneity} for segments where $B_{\rm c} > \Delta\nu_{\rm d}$. The measurements and estimates were used to calculate a weighted mean value of $\Delta\nu_{\rm d}$ for each observing epoch. The anomalous region just before MJD 55500 is primarily over-estimated because of poor dynamic spectra quality because of terrestrial radio interference in some observations at this time. This does not significantly affect the weighted mean for this epoch.}
\label{fig:scint_params}
\end{figure*}

We cut each observation into segments $<12$\,minutes in length (on average they are $\sim 11$\,minutes) and measure the scintillation timescales and bandwidths for each segment (as described in the following section). In this way we can measure the modulation of $\tau_{\rm d}$ across orbital phase for each observation. The decorrelation bandwidth $\Delta\nu_{\rm d}$ measurements and the grouping of observations into the 23 epochs are shown in Figure \ref{fig:scint_params}. 
The epoch to epoch variation represents real changes in the turbulence level of the IISM and is not related to the motion of the pulsar or the Earth. By comparison with the 20 millisecond pulsars monitored by the Parkes Pulsar Timing Array \citep{Keith+13} this variation is quite modest.

The measurements in Figure 2 are from all available archival observations (with $t_{\rm obs}>24$\,mins) during our selected observing span, not just the 126 observations from P361 that we use for modelling. This is because many of the P361 observations have channel bandwidths $B_{\rm c}=0.78$\,MHz, which is greater than $\Delta\nu_{\rm d}$. For these observations the ACF measurement is not useful but we can use the rms flux over the spectrum to estimate the true $\Delta\nu_{\rm d}$. This method is described in Section \ref{sec:inhomogeneity}. The estimates are calibrated by a scaling factor using observations in the same epoch for which $\Delta\nu_{\rm d}>B_c$.

\subsection{Measurement of $\tau_d$ and $\Delta\nu_d$}
\label{sec:measureparams}
We measure the scintillation parameters from $\sim$11\,minute segments of the dynamic spectrum because the scintillation timescale varies rapidly for the relativistic orbit. We characterise the statistics of these segments using the estimated autocovariance function (ACF), $C(\tau,\delta\nu)$. To calculate this we remove the mean, then pad each segment with an equal length of zeroes in both dimensions, perform a 2-D FFT on the zero-padded segment, take the squared magnitude of the result, and perform an inverse 2-D FFT. We then perform a least squares fit of analytical models to
$C(\tau,0)$ and $C(0,\delta\nu)$ to obtain $\tau_{\rm d}$ and $\Delta\nu_{\rm d}$ respectively. First we fit $C(\tau,0)$ with
\begin{align}
\label{eqn:covtime}
C(\tau,0)& =A\exp\left(-\Bigl\lvert \frac{\tau}{\tau_{\rm d}}\Bigr\rvert^{\frac53}\right) \Lambda(\tau,T_{\rm obs}), {\rm for}\ \  \tau > 0 \\
C(0,0)& =W + A.	\nonumber
\end{align}
where $T_{\rm obs}$ is the length of the segment ($\sim$11\,minutes), $\Lambda(\tau,T_{\rm obs})$ is the triangle function of length $\pm T_{\rm obs}$, and $W$ is the variance noise spike. This function is a slight modification to the previous standard of a Gaussian function \citep{Cordes+98}, where the exponent of $5/3$ gives better fit to the shape of Kolmogorov scintillations \citep[e.g.][]{Coles+05, Coles+10, Rickett+14}. This does not bias the estimate of $\tau_{\rm d}$, but slightly reduces its uncertainty. After obtaining $A$, $W$, and $\tau_{\rm d}$, we keep $A$ and $W$ constant because we typically have marginal resolution in frequency and it is easier to isolate the noise spike using the time cut, $C(\tau,0)$. We then fit $C(0,\delta\nu)$ with
\begin{align}
\label{eqn:covfreq}
C(0,\delta\nu)& =A\exp\left(-\Bigl\lvert \frac{\delta\nu}{\Delta\nu_{\rm d} / \ln 2}\Bigr\rvert\right) \Lambda(\delta\nu,B), {\rm for}\ \  \delta\nu > 0 \\
C(0,0)& =W + A	\nonumber
\end{align}
to obtain $\Delta\nu_{\rm d}$, where B is the receiver bandwidth.

This fitting provides straightforward estimators of $\tau_{\rm d}$ and $\Delta\nu_{\rm d}$, which are easily checked manually, but the least squares fit is not optimal because samples of the observed ACF are heavily correlated. We therefore repeat the fit using the same analytical models, but we perform the fit in the Fourier transform domain where we simply transform the autocovariance function and the models.

The 2-D FFT of $C(\tau,\delta\nu)$ is the power spectrum (or secondary spectrum) $P(f_{\rm dop},t_{\rm del})$, where the dimensions are the differential time delay $t_{\rm del}$ and the differential Doppler shift $f_{\rm dop}$ of the interfering waves. To obtain $\tau_{\rm d}$ from this data we first sum $P(f_{\rm dop},t_{\rm del})$ over the $t_{\rm del}$ dimension and divide by the number of samples $N_{\rm del}$ and then we fit the transform of Equation \ref{eqn:covtime}. As before we hold $A$ and $N$ fixed from this fit before obtaining $\Delta\nu_{\rm d}$ by summing $P(f_{\rm dop},t_{\rm del})$ over the $f_{\rm dop}$ dimension, dividing by the number of samples $N_{\rm dop}$, and fitting the resulting power spectrum with the transform of Equation \ref{eqn:covfreq}.

The errors on these average power spectra are independent, but not equal. In fact they are proportional to the average power spectra itself, so we use a weighted least squares fit with the models providing the weights. This approach provides a second (but not independent) estimator for both $\tau_{\rm d}$ and $\Delta\nu_{\rm d}$ for which we believe the measurements and uncertainty estimates are more reliable. We check that the two methods agree within the uncertainty of the second method and review the dynamic spectra if they do not. For this work we use the $\tau_{\rm d}$ and $\Delta\nu_{\rm d}$ measurements from this Fourier-domain method.

\subsection{The effects of inhomogeneity in the IISM}
\label{sec:inhomogeneity}

The primary physical mechanism underlying scintillation is angular scattering by small scale irregularities in electron density that is diffractive in nature. Observations that are sensitive to the spectral exponent of the power-law of density fluctuations have shown a Kolmogorov spectrum truncated at an inner scale \citep[e.g.][]{Armstrong+95, Rickett+09}. Thus it has been assumed that the microstructure in the IISM is turbulent in origin and by default they have been assumed homogeneous. However more recent observations of phenomena such as extreme scattering events (ESEs) \citep[e.g.][]{Fiedler+87, Coles+15} suggest that the turbulence is often inhomogeneous or that inhomogeneous structures that dominate the scattering are often present at some place on the line of sight from the source to the observer.

Furthermore the spatial structure is now often found to be localised (along the line-of-sight) and anisotropic, for example through analysis of parabolic arcs in the two-dimensional fourier transform of the dynamic spectrum that were first discovered by \citet{Stinebring+01}. This has been called the ``secondary spectrum" or the ``delay-Doppler distribution" because its axes are the differential Doppler shift and the differential time delay of the interfering waves that cause the intensity variations \citep[e.g.][]{Walker+04, Cordes+06, Brisken+10}. The parabolic arcs occur when the scattering is dominated by a compact local region on the line of sight. We have analysed the secondary spectra for PSR J1141$-$6545 and found no clear parabolic arcs that could be used provide information on the scattering geometry, likely because of unfavourable sampling. Arcs will be displayed only when dynamic spectra are significantly oversampled in both time and frequency. Thus we have no a priori evidence from our observations for a particular scattering geometry to use in our models of the scintillation velocity. 

The spatial scale of the diffractive scattering, $s_{\rm d}$ is defined as the transverse separation where incident waves have a 1\,radian rms difference in phase. The width of the angular scattering is then $\theta_{\rm d} \approx 1 /\left(k s_{\rm d}\right)$, for incident radiation with wavenumber $k$ \citep{Rickett90}. The radiation received by the observer arrives from a scattering region of diameter $s_{\rm r}=\theta_{\rm d} D_{\rm e}$, where $D_{\rm e}$ is the distance to the scattering screen from the observer. The intensity will also show variations from refraction, with a spatial scale equal to the diameter of this scattering region (the refractive scale). We estimate the size of the scattering region $s_{\rm r}$ from $s_{\rm r}/r_F=\sqrt{\nu/\Delta\nu_{\rm d}}$, where $r_F=\sqrt{D_{\rm e}/k}$ is the ``Fresnel scale," and find $s_{\rm r}\approx 10^8$\,km (AU-scale) for scintillation bandwidth $\Delta\nu_{\rm d}=0.36$\,MHz (mean from Figure \ref{fig:scint_params}) and approximate distance $D_e \sim D=3$\,kpc \citep{Verbiest+12}. Because the projected size of the orbit is $5.6\times 10^5$\,km at this distance, the scattering disk is much larger than the orbital diameter regardless of the location of the scattering screen or a significant error in the pulsar distance. So for a single binary orbit, the line of sight to PSR J1141$-$6545 does not travel outside of the scattering region. Therefore we do not expect the strength of scattering or the anisotropy of the IISM to change during a single orbit. However epoch-to-epoch variation must be expected if the line-of-sight velocity through the scattering region is $\gtrsim 10$\,km\,s$^{-1}$, since in the $\sim 100$\,days between epochs it will traverse the refractive scale $s_{\rm r}$.

We compute the weighted mean and the rms of $\Delta\nu_{\rm d}$ at each epoch, but we find that the apparent value of $\Delta\nu_{\rm d}$ depends on the channel bandwidth $B_{\rm c}$. In a few epochs we have observations with $B_{\rm c} = 0.78$\,MHz and also with $B_{\rm c} = 0.25$\,MHz. In these cases $\Delta\nu_{\rm d}$ should be the same, but the estimator determined from the ACF saturates near $B_{\rm c}$ and typically is close to 0.3\,MHz. Unfortunately many of our observations are made with $B_{\rm c} >  \Delta\nu_{\rm d}$. In this case fitting the ACF does not provide a useful estimate, but we know that the intensity variance $V_I$ will be reduced by a factor of $\Delta\nu_{\rm d} / B_{\rm c}$. When $B_{\rm c}  <  \Delta\nu_{\rm d}$ we know that $V_I = M_I^2$ where $M_I$ is the mean intensity. So when $B_{\rm c} >  \Delta\nu_{\rm d}$ we can use $\Delta\nu_{\rm d} = F_{\rm c}B_{\rm c} V_I / M_I^2$, where $F_{\rm c}$ is a calibration factor that is needed because the method depends on the actual shape of the ACF. We use the epochs for which we have observations with two different values of $B_{\rm c}$ to determine $F_{\rm c}=0.82$. This is shown in Figure \ref{fig:scint_params}, where the direct $\Delta\nu_{\rm d}$ measurements are in dark grey, the scaled estimates from $V_I$ are in light grey, and the resulting weighted mean $\Delta\nu_{\rm d}$ for each epoch is in black. This technique was also used briefly by \citet{Kerr+18}, but they had insufficient data to determine the calibration factor $F_{\rm c}$. We show here for the first time that the method works well enough to detect small changes in bandwidth and generally agrees well with the direct measurement of $\Delta\nu_{\rm d}$ from the ACF.

\begin{figure*}
\centering
\includegraphics[trim={1cm 2cm 1cm 0.5cm},width=\textwidth]{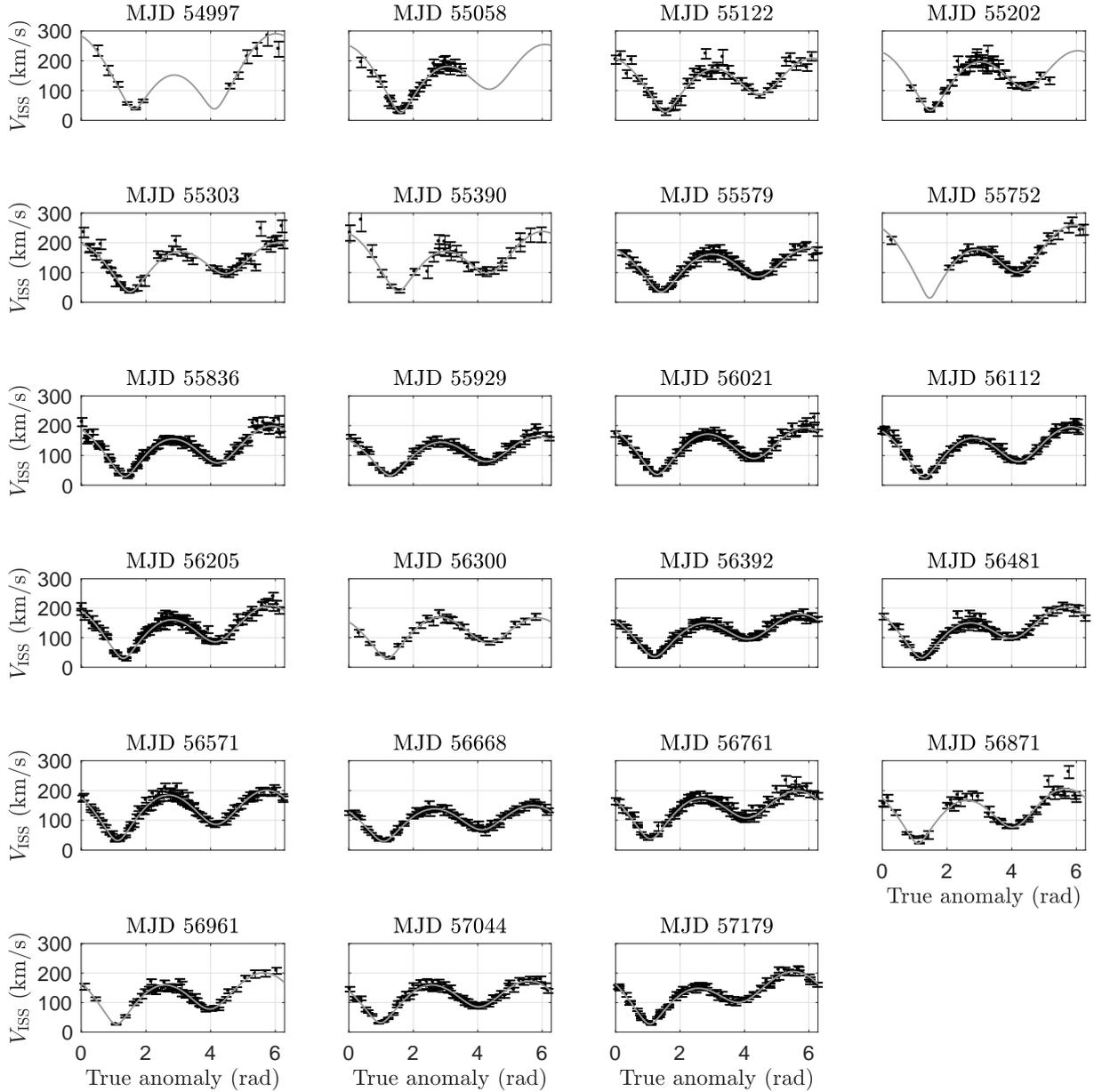}
\caption{Scintillation velocity, $V_{\rm ISS}$ as a function of true anomaly for the 23 epochs of observations shown in Figure \ref{fig:scint_params}. The title of each panel gives the approximate starting date for the first observation in the group. $V_{\rm ISS}$ is defined as the scintillation velocity observed at the Earth for a uniform, Kolmogorov medium along the line-of-sight (Equation \ref{eqn:viss}). The solid line for each panel is the best fit physical model, which is described in Section \ref{sec:shortterm}.}
\label{fig:epochs}
\end{figure*}

\section{The models}
\label{sec:model}
In this section we describe how we use the procedure of \citet{Cordes+98} and a model of the line-of-sight velocity to describe the time variation of $\tau_{\rm d}$ and $\Delta\nu_{\rm d}$. The equations describing the variation of $\tau_{\rm d}$ over a binary orbit are very non-linear in the physical parameters and it is not obvious a priori how many parameters can be estimated from one orbit. However \citet{Rickett+14} showed that the orbital variation of $\tau_{\rm d}$ when the scattering is anisotropic, can be described by 5 harmonic coefficients. In addition one can measure $\Delta\nu_{\rm d}$ which provides an independent degree of freedom. So at best one can model only six parameters which must include: $s_{\rm d}$, the axial ratio of anisotropy $A_{\rm r}$, the angle of anisotropy $\psi$, the distance to screen $s$, and two constant components of the velocity $V_x$, and $V_y$. If $A_{\rm r}$ is small, then  $\omega$ and $i$ can also be measured at each epoch. \citet{Rickett+14} also showed that it was possible to use the shape of the ACF to independently measure the anisotropy, thus providing two more degrees of freedom. However our ACF estimates are not sufficiently accurate to take advantage of this option.

For a long-term analysis we can also use the variation of the Earth's velocity over the year, and the known relativistic advance of the longitude of periastron \citep[from precise pulsar timing;][]{Bhat+08}, to provide additional constraints. Of course one must re-estimate $s_{\rm d}$ from $\Delta\nu_{\rm d}$ at each epoch because the strength of scattering is likely to change on a spatial scale of AU; the estimated size of the scattering region. 

The model of \citet{Cordes+98} is designed to account for variations in the strength of scattering, which will change the $\Delta\nu_{\rm d}$ and consequently $\tau_{\rm d}$. They define a ``scintillation velocity" $V_{\rm ISS}$, which is the ratio of the spatial scale of the diffraction pattern at the observer $s_{\rm d}$, to the temporal scale $\tau_{\rm d}$. This $V_{\rm ISS}$ can be modelled if one knows the distribution and velocity of interstellar plasma along the line-of-sight. We considered two simple models for the plasma distribution: 
\\
\\
\indent {\bf Uniform medium:} A continuous, uniform distribution of plasma with Kolmogorov turbulence  along the line-of-sight. 

\indent {\bf Thin screen:} A single compact ``blob" of plasma with Kolmogorov turbulence, the ``scattering screen", at some position $s$ between the pulsar at $s=0$ and the Earth at $s=1$. 
\\
\\
We also assume that the density irregularities originate from turbulence and are described by a Kolmogorov spectrum. \citet{Cordes+98} derive the spatial scale $s_{\rm d}$ given the observed decorrelation bandwidth $\Delta\nu_{\rm d}$ for both of these models, assuming that the scattering is isotropic. In general the scintillation velocity is given by

\begin{equation}
\label{eqn:viss} 
V_{\rm ISS}=A_{\rm ISS}\frac{\sqrt{D\Delta\nu_{\rm d}}}{f\tau_{\rm d}},
\end{equation}

\noindent where $D$ is the distance to the pulsar in kiloparsecs, $f$ is the observing frequency in GHz, $\Delta\nu_{\rm d}$ is in MHz, and $\tau_{\rm d}$ is in seconds. The factor $A_{\rm ISS}$ depends on the assumed geometry of the scattering medium and on the exponent of the density spectrum. For a uniform medium, $A_{\rm ISS}=2.53\times10^{4}$\,km\,s$^{-1}$, while for a thin screen, $A_{\rm ISS}=2.78\times10^{4}\sqrt{2(1-s)/s}$\,km\,s$^{-1}$ \citep{Cordes+98}. We also include the extension to anisotropic scattering that was first presented by \citet{Coles+05} in both of our models. 

$V_{\rm ISS}$ (Equation \ref{eqn:viss}) is related to the effective transverse line-of-sight velocity $V_{\rm eff}(s)$ through the scattering medium at position $s$, which is a linear combination of the pulsar, Earth, and IISM velocities:

\begin{equation}
\label{eqn:veff1}
V_{\rm eff}(s) = (1 - s)(V_{\rm p} + V_{\mu}) + sV_{\rm E} - V_{\rm IISM}(s),
\end{equation}

\noindent where $V_{\rm p}$, $V_{\mu}$, $V_{\rm E}$, and $V_{\rm IISM}$ are the velocities from the pulsar's orbit, the pulsar proper motion, the Earth, and the IISM respectively. Each is relative to the Solar system barycentre, and the IISM velocity can vary as a function of distance along the line-of-sight, $s$, for extended scattering. In Appendix A we describe the model for $V_{\rm eff}$, which includes the definitions for the pulsar orbital velocity, the orientation of the orbit in celestial coordinates, and the extension to anisotropic scattering developed by \citet{Coles+05}.

In addition to these models that we use for the long-term scintillation, we also consider the short-term model of \citet{Ord+02a} for each of the 23 epochs of observations in Figure \ref{fig:epochs}. This physical model necessarily assumes isotropic scattering, but we also fit the harmonic coefficient model of \citet{Rickett+14}, which is useful for understanding sources of noise in our long-term analysis. Both of these models are summarised in Section \ref{sec:shortterm} with results given in Section \ref{sec:singleepochs}.

Each scintillation velocity model includes a single constant component, which (from Equation \ref{eqn:veff1}) is $V_C = (1-s)V_\mu - V_{\rm{IISM}}$ for a thin screen (for a uniform medium there is an appropriate integration over $s$ with $V_{\rm{IISM}}(s)$; Appendix A, Equation \ref{eqn:viss_int}). The proper motion of PSR J1141$-$6545 is not currently known from pulsar timing, however we expect that the transverse velocity of the pulsar system ($V_{\mu}$) is larger than any IISM velocity ($V_{\rm{IISM}}$), since the latter may be of order $\sim 10$\,km\,s$^{-1}$. Therefore if we parameterise $V_C$ with only $V_\mu$ in our models (i.e. by setting $V_{\rm{IISM}}=0$ in Equation \ref{eqn:veff1}), the measurement will include a contamination from any non-zero $V_{\rm IISM}$. So when interpreting our measurement of $V_\mu$ as the proper motion, we would implicitly make the assumption that the velocity of the scattering screen is small compared with $V_\mu$, i.e. $V_{\rm{IISM}}<<(1-s)V_\mu$. To account for this, we include a conservative $10$\,km\,s$^{-1}$ factor added in quadrature to the measurement uncertainty of each component of $V_\mu$.   

The parameters in our long-term $V_{\rm{eff}}$ model that are unknown from pulsar timing of PSR J1141$-$6545 are $s$, $i$, $\Omega$, $v_{\mu,\parallel}$, $v_{\mu,\perp}$, $R$, and $\psi$. The subscripts $\parallel$ and $\perp$ denote coordinates parallel and perpendicular to the line of nodes of the pulsar's orbit (Appendix A). We also require a scaling factor $\kappa$ to account for any systematic errors in the calculation of $s_{\rm d}$ from $\Delta\nu_{\rm d}$. The constraints from the known Earth's velocity and $\dot{\omega}$ provide the additional degrees of freedom required to uniquely determine these parameters in a long-term analysis. For the uniform medium model we have one fewer parameter because we integrate over $s$ (Equation \ref{eqn:viss_int} in Appendix A). For the case of isotropic scattering, which we consider for both the uniform medium model and the thin screen model, we have $R=0$ (reducing the model to Equation \ref{eqn:veff_isotropic} in Appendix A, scaled by the factor $\kappa$). However, we also found in Section \ref{sec:inhomogeneity} that it is necessary to re-estimate the spatial scale from $\Delta\nu_{\rm d}$ at each epoch because of AU-scale inhomogeneities in the IISM. 

The calculation of the spatial scale from these $\Delta\nu_{\rm d}$ measurements is imperfect, and \citet{Rickett+14} found some disagreement between $\Delta\nu_{\rm d}$ and the spatial scale (by analysing the harmonic coefficients described in the following section). To account for this, and some other correlated time-variability in properties such as the anisotropy and IISM velocity, we chose to use a scaling factor at each epoch. To do this, we scaled the data for each epoch to the mean $V_{\rm ISS}$, then fitted for the mean scaling factor $\kappa$ between the model and scaled data. Ideally $\kappa$ would then represent the major systematic biases in the model, such as from an inaccurate assumed pulsar distance, while the scaling at each epoch accounts for time variations in $s_{\rm d}$ and perhaps some of the variation due to changing anisotropy and IISM velocity.

\subsection{Short-term models of earlier work}
\label{sec:shortterm}

For the first scintillation analysis of PSR J1141$-$6545, \citet{Ord+02a} modelled the dynamic spectrum for a single 10\,hr observation. From this dynamic spectrum they derived the $V_{\rm ISS}$ of \citet{Cordes+98} (Equation \ref{eqn:viss}), and modelled this using only the pulsar's orbital velocity components ($v_{\mu,\parallel}$ and $v_{\mu,\perp}$; Appendix A, Equation \ref{eqn:vcomponents}), with a scaling factor $\kappa_u$. They assumed that the scattering was isotropic and the velocity model was then given by

\begin{equation}
V_{\rm model}=\kappa_u\sqrt{v_{{\rm p},\parallel}^2 + v_{{\rm p},\perp}^2},
\end{equation}

\noindent with $\kappa_u$, $i$, $\omega$, $v_{\rm C,\parallel}$, and $v_{\rm C,\perp}$ being the five fitted parameters. We use $v_{\rm C,\parallel}$ and $v_{\rm C,\perp}$ here instead of $v_{\mu,\parallel}$ and $v_{\mu,\perp}$ because they will include a contribution from the Earth's velocity, and as a result will show annual variations with time. The parameter $\kappa_u$ can be used to absorb any errors in the calculation of $V_{\rm ISS}$, such as from an incorrect pulsar distance, which is what we intend for $\kappa$ of the long-term models (Equation \ref{eqn:veff_anisotropic}). However, since the pulsar velocity in this model is not scaled by a screen distance $s$ because the Earth's velocity is not included to constrain it, one could also interpret $\kappa_u$ in this case as a scale factor for $A_{\rm ISS}$. In this interpretation, $\kappa_u$ is also related to the screen distance $s$ and can account for different scattering geometries provided they are isotropic. 

We repeat this analysis of \citet{Ord+02a} as described above for each of the 23 epochs of observations and we refer to this as the single-epoch ``physical model". We present the results from this model in Section \ref{sec:singleepochs} and show that this approach can also be used to measure the advance of periastron $\dot{\omega}$.

Like PSR J1141$-$6545, which we analyse here, the double pulsar PSR J0737$-$3039A shows orbital modulation of the scintillation timescale, which was first measured and modelled by \citet{Ransom+04}. The analysis was extended to include anisotropy in the IISM by \citet{Coles+05} and further extended to include the Earth's velocity by \citet{Rickett+14}, who also showed that the modulation of $1/\tau_{\rm d}^2$ with orbital phase can be modelled as the sum of five harmonics 

\begin{equation}
\label{eqn:harmcoeffs}
\frac{1}{\tau_{\rm d}(\phi)^2} = K_0 + K_S\sin\phi + K_C\cos\phi + K_{S2}\sin(2\phi) + K_{C2}\cos(2\phi),
\end{equation} 

\noindent where the harmonic coefficients contain all of the information on the diffractive interstellar scintillation available in the data. The relationship between these coefficients and the physical parameters of the scattering and velocity are given in Equation 10 of \citet{Rickett+14} and are reproduced in Appendix A (Equation \ref{eqn:normharms})

It is useful to note that each of the coefficients are inversely proportional to $s_{\rm d}^2$, and that $K_{S2}$ and $K_{C2}$ are constant with time for a constant anisotropy. \citet{Rickett+14} used these facts to normalise the coefficients by $K_{C2}$, which corrects each one for changes to the spatial scale $s_{\rm d}$ with time and with observing frequency. This allowed them to model observations at multiple observing frequencies simultaneously. 

\begin{figure}
\includegraphics[width=0.5\textwidth]{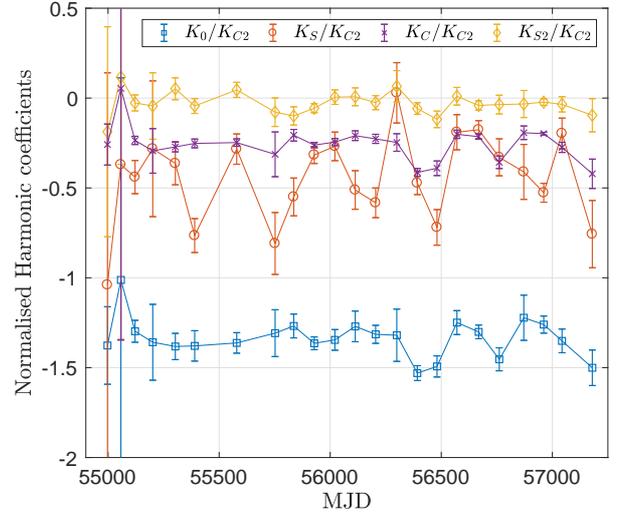}
\caption{The normalised harmonic coefficients derived from a fit to $V_{\rm ISS}(\phi)$ at each observing epoch, as described in Section \ref{sec:shortterm}. A colour version of this figure is available through the online journal.}
\label{fig:harmcoeffs}
\end{figure}

We have measured these normalised harmonic coefficients ($k_0 = K_0/K_{C2}$, $k_S = K_S/K_{C2}$, $k_C = K_C/K_{C2}$, and  $k_{S2} = K_{S2}/K_{C2}$) for each epoch of PSR J1141$-$6545 observations. The results are shown in Figure \ref{fig:harmcoeffs}. It is worth noting that in the case of isotropic scattering (with $a=b=1$ and $c=0$), $k_S$ only changes with time because of $V_{\rm C,\parallel}$ (ignoring $\dot{\omega}$), $k_{C}$ changes with $V_{\rm C,\perp}$, and $k_{S2}=0$. While we do find that $k_{S2}$ is consistent with zero on average (Figure \ref{fig:harmcoeffs}), there is some variation that is correlated with variations in the other normalised parameters. Since $k_{S2}$ is constant with orbital phase (for a constant anisotropy over the orbit), this indicates epoch-to-epoch variation to the anisotropy, which is an unmodelled source of noise in our data (as well as most scintillation observations). Because $k_{S2}=0$ on average, we can also see that our long-term model must be consistent with $c=0$, and thus should indicate that the scattering is isotropic, with $R=0$, or anisotropic with orientation near $\psi=0^\circ$ or $\psi=90^\circ$.

The harmonic coefficient model for each epoch was found with a linear least squares fit, while the physical model of \citet{Ord+02a} was fitted with nonlinear regression (using the \textsc{lsqnonlin} function in \textsc{Matlab}) because this was a convenient way to propagate uncertainties to the physical parameters. For the long-term models we use Bayesian methods for determining parameter uncertainties and performing model selection, as described in the following Section.

\section{Bayesian inference and model selection}
\label{sec:fitandcompare}

Bayesian methods have recently become popular for analysing astrophysical datasets because they can provide a robust means for parameter estimation, model selection, and visualisation of parameter correlations. An introduction to Bayesian inference, with examples taken from gravitational-wave astronomy, is given in \citet{Thrane+18}. In brief, we aim to estimate the posterior probability distribution $P(\theta|d)$ for our set of model parameters $\theta$, given our data $d$, using Bayes theorem:
\begin{equation}
P(\theta|d) = \frac{P(d|\theta)P(\theta)}{Z},
\end{equation}
\noindent where $P(d|\theta)$ is the likelihood function of the data given the set of parameters of a model, $P(\theta)$ is the prior probability distribution for the parameters, and $Z\equiv P(d)=\int P(d|\theta)P(\theta)d\theta$ is the fully-marginalised likelihood function or ``evidence". For a single model, the evidence $Z$ is a normalisation factor that is often ignored, but it becomes meaningful when comparing evidences for multiple models. For each model of scattering geometry, we sample $P(\theta|d)$ and estimate $Z$ with the sampler of \citet{Veitch+10}, implemented in MATLAB by \citet{Pitkin+12}, with 1000 initial sample points and a multivariate Gaussian likelihood function. 

We initially used a uniform prior for each parameter across a physically-motivated bound region: $\Omega\in [0,2\pi)$, $\cos{i}\in [-1,1]$, $v_{\mu,\parallel},v_{\mu,\perp} \in [-2000,2000]$\,km\,s$^{-1}$, $R \in [0,1)$, $\psi\in [0,\pi)$, $\kappa\in (0,3]$, and $s\in (0,1)$. However, for the anisotropic uniform medium model, we found that the resulting solution was non-physical (because the measured inclination angle $i= 48.6\pm1.9^\circ$ is ruled out by timing analysis), suggesting that the model is incorrect. To test this, we also consider each of these models with a Gaussian prior probability distribution on $i$ with mean and standard deviation $73\pm 3^\circ$\footnote{Although each model with a uniform $\cos{i}$ prior returned a unique solution with $i<90^\circ$, we also considered the opposite sense for $i$, with a Gaussian prior of $107\pm3^\circ$ to quantify the strength of evidence for the measurement of the sense.}. This prior corresponds to the 2$\sigma$ range derived using GR in \citet{Bhat+08}. We choose this 2$\sigma$ range because the estimate in \citet{Bhat+08} did not account for correlations between $i$ and other parameters in the timing model, because it was not a true measurement of the Shapiro delay.
 
 We also found that there are significant correlations between $i$, $v_{\mu,\perp}$, and $R$ \citep[Figure \ref{fig:posterior_i73}, which has also been reported previously, e.g.][]{Coles+05,Rickett+14}. Using the Gaussian prior distribution on $i$ allows us to constrain the models to something physical, given that the anisotropy is completely unconstrained a priori. In addition $v_{\mu,\perp}$ and $R$ likely have some unmodelled time-dependence, which may contribute to a bias in these parameters and correlated parameters such as $i$. Using this Gaussian prior will therefore give a more reliable estimate of the anisotropy and pulsar velocity.

\subsection{Bayes factors}
\label{sec:bayesfactors}
For the purpose of model selection, we choose the null hypothesis to be isotropic scattering by a thin screen with a uniform prior for $\cos{i}$. This can be considered as the standard model for scintillation velocity analysis, but this is often out of necessity because of degrees-of-freedom constraints.  We use the Bayes factor $K$ to quantify the evidence for (or against) our test models. This Bayes factor is the ratio of model evidences, which we choose to quote as the difference between log-evidences
\begin{equation}
\label{eqn:bayesfactor}
    \log{K} = \log{Z_h} - \log{Z_0},  
\end{equation}
\noindent where $Z_0$ and $Z_h$ are the evidence for the null hypothesis and test model respectively.

There are numerous suggestions for how to interpret the strength of evidence for a model, given a Bayes factor \citep[e.g.][]{Jeffreys98, Kass+95}. $|\log{K}|>8$ is commonly used as a significance threshold, while $|\log{K}|>20$ can be considered ``strong" evidence for one model over the other, with the sign indicating which model is favoured \citep{Kass+95}. However, for this work we find that the magnitude of Bayes factors are much greater than these contentious thresholds. 

In Table \ref{tab:compmodel} we show the $\log{K}$ for each long-term model with respect to the isotropic thin screen. In the top half of this table we show the values obtained with a uniform prior for $\cos{i}$. This shows that the isotropic uniform medium is decisively ruled out, while the anisotropic thin screen and anisotropic uniform medium are strongly preferred by the data. Although there is strong evidence for the anisotropic thin screen over the anisotropic uniform medium, with $\log{K}=28.3$, we also reconsidered these models with a Gaussian prior on $i$ to help constrain these models to physical solutions. The Bayes factors for these models are in the lower half of Table \ref{tab:compmodel} and show that only an anisotropic thin screen is favoured over the isotropic screen. 

Finally, to quantify the evidence for our measurement of the sense of the inclination angle, we also considered Gaussian priors of $107\pm 3^\circ$ with each model. Each model strongly favours $i<90^\circ$, for example comparing the two priors for the anisotropic thin screen we have $\log{K}=178.6$ in favour of $73\pm 3^\circ$.

The decisive evidence for anisotropic scattering reflects the fact that a long-term analysis such as this is highly sensitive to changes in $\tau_{\rm d}$ as a function of the position angle of $V_{\rm eff}$. This is particularly true because the annual variation is completely known except for a scale factor, and this is also the reason for such strong evidence for the thin screen model and inclination $<90^\circ$. This highlights the importance of annual variations for modelling scintillation velocities.

\begin{table}
\centering
\caption{Model comparison with Bayes factors ($\log{K}$, Equation \ref{eqn:bayesfactor}) relative to the standard model of an isotropic thin screen.}
\begin{tabular}{@{}cc@{}} 
\hline
\hline
Geometry &  $\log{K}$ \\
 \hline
 \multicolumn{2}{c}{Uniform $\cos{i}$ prior} \\
 \hline
Isotropic screen  & 0 \\
Isotropic uniform & $-$292.0  \\ 
Anisotropic screen  & \textbf{82.9} \\
Anisotropic uniform  & 54.6  \\
\hline
 \multicolumn{2}{c}{Gaussian $i$ prior: $73\pm 3^\circ$} \\
 \hline
Isotropic screen & $-$4.6 \\
Isotropic uniform  & $-$426.5 \\
Anisotropic screen & \textbf{82.1} \\
Anisotropic uniform & $-$64.0 \\
 \hline
 \label{tab:compmodel}
\end{tabular}
\end{table}

\section{Results}
\label{sec:results} 

Here we present and compare the various scintillation models for PSR J1141$-$6545. We first fitted each of the 23 epochs separately with a sum of five harmonics of orbital phase, which are described in \citet{Rickett+14} and in Section \ref{sec:shortterm}. We then fit a physical model for isotropic scattering to each epoch (Section \ref{sec:shortterm}), which provided measurements of $\omega$ and $V_{\rm C}$ over time to clearly show that the relativistic advance of periastron and the modulation from the Earth's velocity can be recovered in the data. These models are presented below in Section \ref{sec:singleepochs}.

We then held the periastron advance fixed at the precisely-measured value from pulsar timing, $\dot{\omega}=5.3096$\,$^\circ{\rm yr}^{-1}$ \citep{Bhat+08}, and calculated the components of the Earth's velocity ($V_{\rm E}$) transverse to the line-of-sight of PSR J1141$-$6545 for each observation. We combined the 23 epochs of data and fit several long-term scintillation models, using the additional degrees of freedom provided by $\dot{\omega}$ and $V_{\rm E}$ to help constrain additional parameters including the scattering anisotropy ($R$ and $\psi$) and the longitude of the ascending node $\Omega$. We fit these long-term velocity models for two distributions of Kolmogorov turbulent plasma along the line-of sight: A uniform distribution, and a thin screen. The uniform medium models are ruled out with high significance, as is purely isotropic scattering (see previous Section). We present our long-term anisotropic thin screen model in Section \ref{sec:anisotropic} for two prior probability distributions on the inclination angle.

Finally, we have included in Appendix B the links to access the data (raw observations and processed dynamic spectra) used for this work, as well as the MATLAB code used for the analysis for the purpose of reproducibility, including the scripts used to measure scintillation parameters from the dynamic spectra and fit the scintillation velocity as described in the previous section.

\subsection{Individual epochs}
\label{sec:singleepochs}

For each epoch of observations shown in Figure \ref{fig:epochs}, we fit the $V_{\rm ISS}$ of Equation \ref{eqn:viss} as a function of orbital phase $\phi$, with a sum of five harmonics (Section \ref{sec:shortterm}). The resulting normalised harmonic coefficients are given in Figure \ref{fig:harmcoeffs} as a function of time. In the case of PSR J0737$-$3039A, the inclination of the orbit is essentially edge-on, which means that $k_C$ and $k_{S2}$ are almost zero \citep{Rickett+14}. This is not true for PSR J1141$-$6545, but we do still find that the mean of $k_{S2}$ is close to zero. This suggests that the scattering may be nearly isotropic since $k_{S2}$ is independent of the changing pulsar and Earth velocities. There is however some time variability in $k_{S2}$, which is correlated with some other normalised harmonic coefficients. This may be due to time variability in the anisotropy, which one would expect even from random realisations of a truly isotropic medium. This time variability is a source of noise for the long-term scintillation models. This may be the primary reason for a large reduced chi-squared value ($\chi_r^2 \sim 2.3$) for the long-term models (e.g. Figure \ref{fig:viss}).

Considering the evidence for isotropic scattering is important because the single-epoch physical model \citep{Ord+02a} depends on it. With only five degrees of freedom available from a single epoch of observations, accurate measurement of the pulsar proper motion and inclination angle can only be made for isotropic scattering \citep{Coles+05}. We therefore fit this physical model to each of the epochs under the assumption of isotropic scattering to obtain a time series for each of the physical parameters described in Section \ref{sec:shortterm}. 

There are four possible solutions for each epoch, arising from a known degeneracy between the proper motion and inclination angle \citep{Lyne84}. This produces a more edge-on solution with higher proper motion and a more face-on solution with lower proper motion, and their corresponding pairs with an opposite ``sense" of inclination about $i=90^\circ$. \citet{Ord+02a} considered only the $i<90^\circ$ solutions and determined the more edge-on solution to be favorable using the implied pulsar mass. For each of our fits we also take the $i<90^\circ$ solutions, and we show in Sections \ref{sec:bayesfactors} and \ref{sec:anisotropic} that this is physical. The measurement of $\omega$ is unaffected by these degeneracies. 
\begin{figure}
\includegraphics[width=0.5\textwidth]{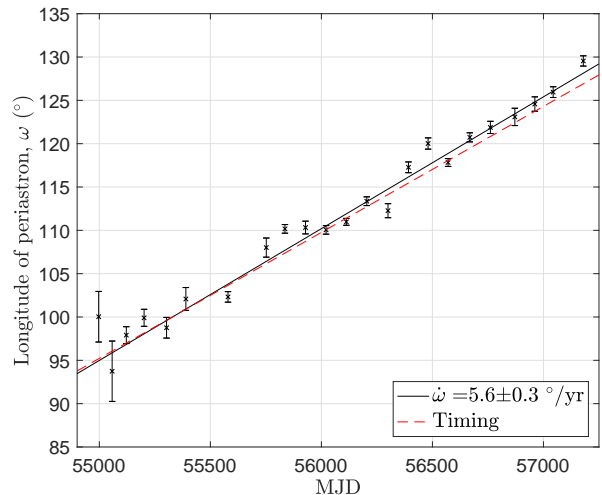}
\caption{The longitude of periastron, $\omega$ as a function of time, measured independently at each of the 23 epochs shown in Figure \ref{fig:epochs}. The solid line is a weighted best fit, where the gradient, $\dot{\omega}=5.6\pm0.3$\,$^\circ$\,yr$^{-1}$, is the advance of periastron and is consistent with the measurement from pulsar timing (dashed line).}
\label{fig:omdot}
\end{figure}
The time series of our fitted $\omega$ values is shown in Figure \ref{fig:omdot}. The clear gradient is the relativistic advance of periastron, $\dot{\omega}=5.6\pm0.3$\,$^\circ/{\rm yr}$, and is close to the measurement of pulsar timing $\dot{\omega}=5.3096\pm0.0004$\,$^\circ/{\rm yr}$ \citep{Bhat+08}. We find that $\kappa_u$ and $i$ are constant with time with the exception of random variations similar to those seen in the harmonic coefficients (Figure \ref{fig:harmcoeffs}). The locus of the two solutions in inclination angle from these degenerate solutions are $i=80.1\pm1.1^\circ$ and $i=71\pm3^\circ$, and the magnitudes of their corresponding velocities are $76\pm3$\,km\,s$^{-1}$ and $44\pm 2$\,km\,s$^{-1}$ respectively. These inclination angles are both consistent with the previous scintillation measurement of $76\pm2.5^\circ$, to within about 1$\sigma$ of both measurements. 

\begin{figure}
\includegraphics[width=0.5\textwidth]{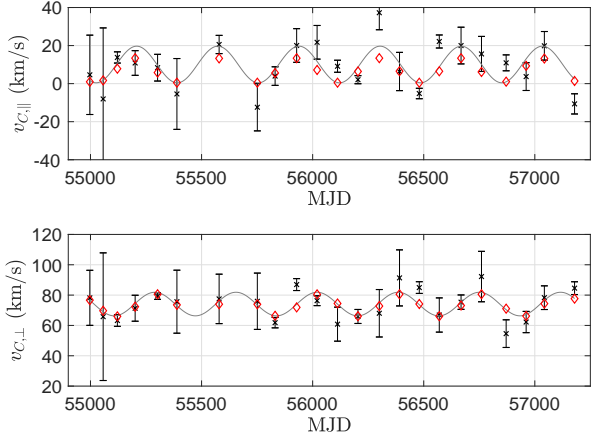}
\caption{Components of the constant (with orbital phase) transverse velocity parallel ($v_{\rm C,\parallel}$, top panel) and perpendicular ($v_{\rm C,\perp}$, bottom panel) to the line of nodes for each observing epoch. The measured velocity is a scaled combination of the Earth and IISM velocities and the pulsar proper motion. The Earth's contribution is apparent from the clear annual modulation. The solid line is a weighted best fit annual sine wave to each of the time series. The diamonds are the calculated Earth's velocity at each epoch, scaled down by a screen distance $s=0.36\pm0.07$, and rotated from celestial coordinates by $\Omega=25\pm10^\circ$, according to the fit.}
\label{fig:ordVelocity}
\end{figure}

The time-series of $V_C$ components for the more edge-on solution is shown in Figure \ref{fig:ordVelocity} to demonstrate the clear annual variation, however we do not comment on which solution is physical because these parameters are highly correlated with $R$ and are therefore both biassed under the assumption of isotropic scattering. The reduced chi-squared value for these solutions (across all epochs, with total number of parameters $m=115$; five per epoch), is $\sim \chi^2_r=1.5$, suggesting that either the measurement errors for scintillation parameters are underestimated or there is excess noise in the data. The $\chi^2_r$ for individual epochs with observations spanning several days is generally higher than the few epochs with a single day, which is usually close to unity. It is therefore likely that the cause is small random changes in the scattering on a timescale of $\sim$days, such as variations to the anisotropy (which we would expect even from isotropic scattering).

\subsection{Long-term model: Anisotropic thin screen}
\label{sec:anisotropic}

The annual variations of $V_{\rm ISS}$, which are clear in the single-epoch models shown in Figure \ref{fig:ordVelocity}, are due to the known $V_{\rm E}$ and can be included in our long-term model. This provides additional constraints that break the usual degeneracies between $i$ and $v_{\mu,\perp}$, and in the sense of $i$. It also provides the additional degrees of freedom required to measure the scattering anisotropy.

\begin{table}
\centering
\caption{Measured and derived parameters for the long-term scintillation model of an anisotropic thin screen, using two different priors for the inclination angle $i$. The Gaussian prior had a mean and standard deviation of $73^\circ$ and $3^\circ$ respectively, corresponding to the 2$\sigma$ confidence region of \citep{Bhat+08}. The value given in brackets are the uncertainty on the last quoted decimal place. The uncertainties for our measurements of $v_{\mu,\parallel}$ and $v_{\mu,\perp}$ have been inflated with $10$\,km\,s$^{-1}$ added in quadrature to account for possible contributions from an unknown $V_{\rm{IISM}}$ of approximately this magnitude.}
\begin{tabular}{@{}lcc@{}} 
\hline
\hline
\multicolumn{3}{c}{Measured parameters} \\
\hline
& \multicolumn{2}{c}{Inclination angle prior} \\
& Uniform $\cos{i}$ & Gaussian $73\pm3^\circ$\\
 \hline
$\kappa$  				& $0.432(17)$				& $0.496(14)$ 			\\
$s$   										& $0.299(9)$			& $0.276(8)$ 			\\
$\Omega$ ($^\circ$)  				& $27.1(18)$					& $24.8(18)$		 			\\
$i$ ($^\circ$)			& $61.3(20)$ 					& $68.6(1.1)$			 			\\
$v_{\mu,\parallel}$ (km\,s$^{-1}$)		& $21(11)$				& $19(11)$	 			\\
$v_{\mu,\perp}$ (km\,s$^{-1}$)			& $242(23)$				& $173(15)$	 			\\
$\psi$ ($^\circ$)  					& $87.8(3)$				& $86.8(4)$	 			\\
$R$  										& $0.77(3)$				& $0.64(3)$	 			\\
\hline
\hline
\multicolumn{3}{c}{Derived parameters} \\
\hline
$A_r$									&		$2.80(18)$				&		$2.14(11)$			\\
$M_{\rm p}$ ($M_\odot$)		&		$1.18(2)$			&		$1.242(8)$				\\
$M_{\rm c}$ ($M_\odot$)		&		$1.11(3)$			&		$1.047(8)$				\\
$D$ (kpc)								&		$12^{+4}_{-3}$	&		$10^{+4}_{-3}$	\\
$\mu_\alpha\cos{\delta}$ (mas\,yr$^{-1}$)			&		$3.3(10)$			&		$2.9(10)$			\\
$\mu_\delta$ (mas\,yr$^{-1}$)				&		$2.1(6)$			&		$1.8(6)$				\\
\hline
\label{tab:bestparams}
\end{tabular}
\end{table}

Any anisotropy cannot be accounted for by measuring a single epoch of $V_{\rm ISS}$ alone because there are too few degrees of freedom available for the additional two parameters. Instead, \citet{Coles+05} was able to use the correlated scintillations of both pulsars in the double pulsar system at their apparent closest approach (when the magnetosphere of pulsar B eclipses pulsar A) to produce a spatial correlation pattern that revealed the anisotropy. Later, \citet{Rickett+14} was able to use the annual variations in the harmonic coefficients for PSR J0737$-$3039A to measure the anisotropy and determine the sense of the inclination angle by comparing the two models with inclinations fixed by the $\sin{i}$ measurement from pulsar timing. They also showed that the anisotropy and IISM velocity (because the proper motion velocity was known from timing) could be measured from a model of the individual ACFs alone, but that the measurements differed from the average value obtained with long-term fitting, suggesting some weakness in the model and difficultly in confidently determining the anisotropy. The annual variation approach is equivalent to the long-term physical models we use here, but we do not have sufficient frequency resolution to use the ACFs of PSR J1141$-$6545 for an independent anisotropy estimation. Fortunately though, our data are sensitive to the changing spatial scale for different orientations of the rotating velocity vector, and we have been able to rule-out a purely isotropic model with high significance.  
\begin{figure*}
\includegraphics[width=0.9\textwidth]{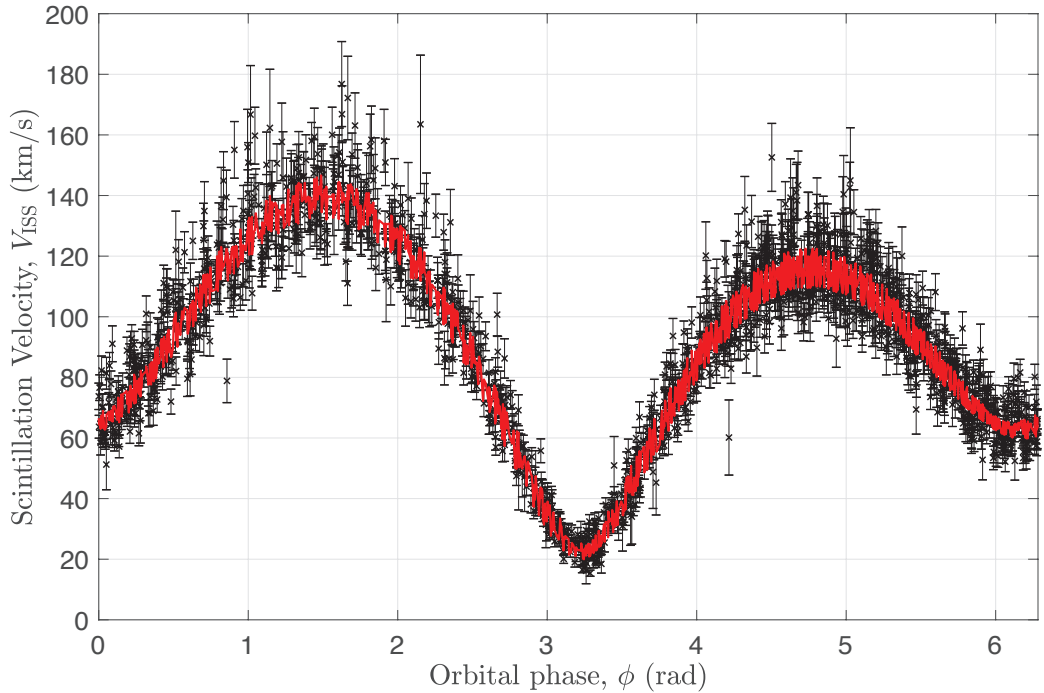}
\caption{Scintillation velocity, $V_{\rm ISS}$ as a function of orbital phase, $\phi$ for $\sim$6\,years of PSR J1141$-$6545 observations at a frequency of $\sim1400$\,MHz. The orbital phase was calculated using the $\omega$ and $\dot{\omega}$ values measured in the pulsar timing model. $V_{\rm ISS}$ in this case is defined at the scattering screen (IISM frame), because the distance to the screen was a parameter in the model. The best-fit anisotropic thin screen model is shown as the red line, and the apparent white-noise in the model is due to the out-of-phase Earth's velocity. The data in each epoch have been scaled to the mean $V_{\rm ISS}$ and the model is scaled by $\kappa$ to match. A colour version of this figure is available in the online version of the journal.}
\label{fig:viss}
\end{figure*}

\begin{figure*}
\includegraphics[trim={4cm 2cm 4cm 2cm},clip,width=\textwidth]{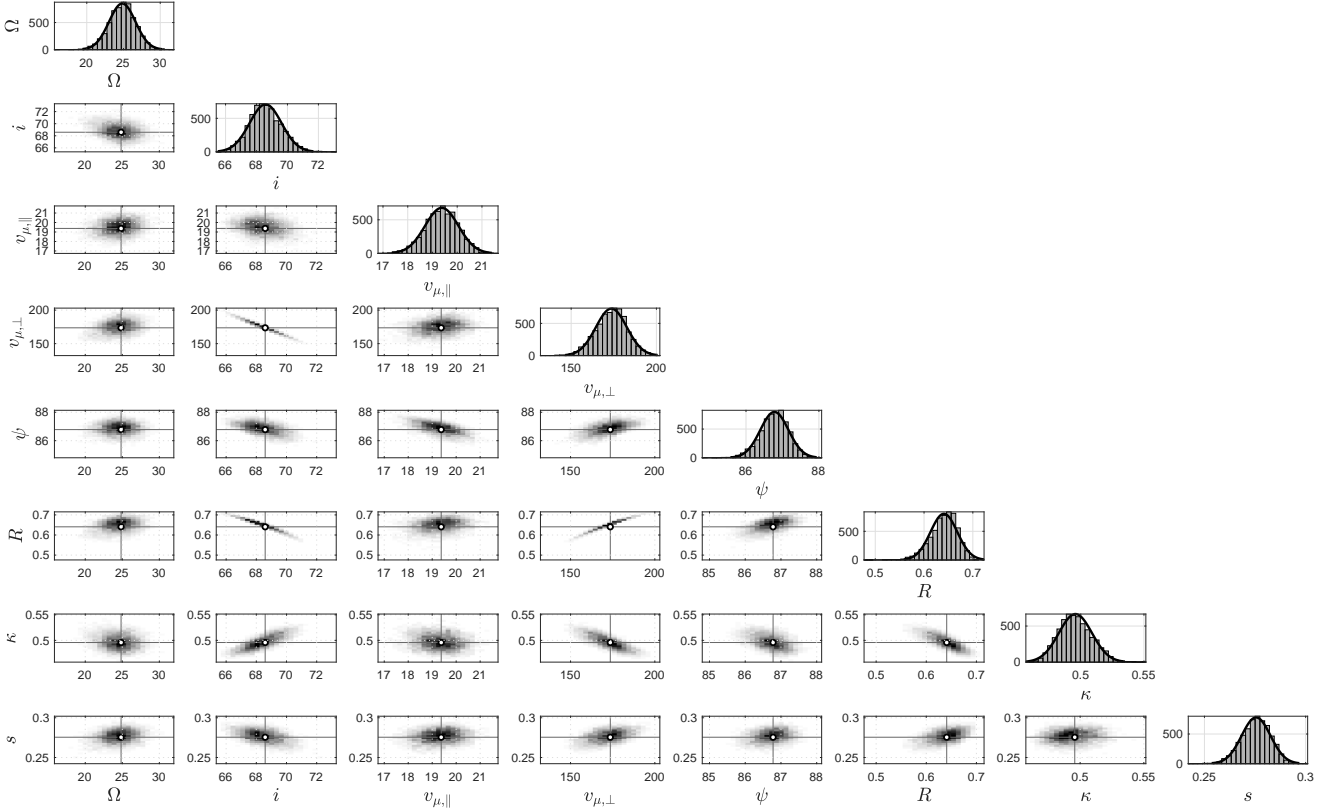}
\caption{One- and two-dimensional posterior probability distributions for parameters of the anisotropic thin screen model, using a Gaussian prior for the inclination angle $i = 73 \pm 3 ^\circ$ \citep[$2\sigma$ result from][]{Bhat+08}. White dots in each two-dimensional distribution mark the mean for the corresponding parameters. The angles $\Omega$, $i$, and $\psi$ are shown with units of degrees, while $v_{\mu,\parallel}$ and $v_{\mu,\perp}$ are in km\,s$^{-1}$. The mean and standard deviation for each parameter is given in Table \ref{tab:bestparams}, with additional derived parameters. Since $v_{\mu,\parallel}$ and $v_{\mu,\perp}$ will include any IISM velocity, we inflate the uncertainty quoted in  Table \ref{tab:bestparams} by adding $10$\,km\,s$^{-1}$ in quadrature with the standard deviation from this posterior.}
\label{fig:posterior_i73}
\end{figure*}

Our best-fit long-term model, which has an anisotropic thin screen and includes a Gaussian prior on inclination angle, is shown as $V_{\rm ISS}(\phi )$ in Figure \ref{fig:viss}. The model parameters are given in the second column of Table \ref{tab:bestparams}. The full posterior probability distributions for these parameters are shown in Figure \ref{fig:posterior_i73}. Each observation epoch has been independently scaled to the mean $V_{\rm ISS}$ in Figure \ref{fig:viss} as described in Section \ref{sec:model}, and the apparent white noise in the model is actually the out-of-phase variation due to the Earth's velocity. For this model we find $i=68.6\pm1.1^\circ$, which is more than $1\sigma$ inconsistent with the prior, suggesting a strong preference in the data for a lower inclination angle. The anisotropy for this model is $R=0.64\pm0.03$, which corresponds to an axial ratio $A_r=2.14\pm 0.11$. This is comparable in magnitude to what has been observed for scattering towards other objects \citep[e.g.][]{Coles+05, Frail+94}.

If the scattering were truly isotropic, we would expect that each observation samples a single realisation of this, and thus would randomly appear slightly anisotropic, with an rms of $R_{\rm rms}\sim0.7(s_{\rm d}/s_{\rm r})^{-1/6}$ \citep{Romani+86,Rickett+14}. In our case $R_{\rm rms}\sim0.18$, and from 23 epochs we would expect to observe $R_{\rm rms}/\sqrt{23}\sim0.04$. The measurement is significantly larger than this, which suggests that there is significant anisotropy in the scattering. We find $\psi = 86.8\pm0.4^\circ$, which is a rather close alignment with the pulsar orbit. Given that the IISM is independent of the pulsar, $\psi$ must be uniformly distributed over 180$^\circ$. So the probability of such alignment is only 3.6\%. However including the anisotropy greatly improves the fit and the accuracy of all the derived parameters, so we have to assign the alignment to chance.

Anisotropy of this magnitude, in the regime of strong scattering, is expected to produce a set of reversed sub-arclets distributed along a primary arc in the secondary spectra \citep[e.g.][]{Walker+04, Cordes+06, Brisken+10}. However, without fine sampling in the dynamic spectra, this collection of arclets would together appear as broad unfocussed power and may not have a clear curvature. Indeed our analysis of the secondary spectra failed to show any clear scintillation arcs, which may be due to a combination of the sampling characteristics of the dynamic spectra and the broadening of the arcs through reverse sub-arclets. Future observations and studies of PSR J1141$-$6545 could be designed to maximise the detail in the ACFs and secondary spectra to allow for an independent measurements of the anisotropy and its time variability, which could then be used in these long-term $V_{\rm ISS}$ models to improve the $i$ and $v_{\mu,\perp}$ measurements.

The reduced chi-squared value for the best models is $\chi_r^2\sim2.3$, and (as mentioned for the single-epoch models as well) this is likely high because of model errors, such as our assumption of a time-stationary anisotropy and IISM component in $v_{\mu}$.  Indeed the normalised harmonic coefficients in Figure \ref{fig:harmcoeffs} show some time variability that would not be explained by the Earth's motion or by changes to the strength of scattering that would be taken into account by scaling of each epoch. This random time-variability is likely to be our largest source of noise and the major reason for a large $\chi_r^2$, but there may also be some contribution from measurement errors on the scintillation parameters for example. Obtaining independent constraints on the anisotropy and IISM velocity from modelling high-resolution ACFs directly \citep[e.g.][]{Rickett+14} is the best chance for future work to improve the quality of the fit.

\section{Discussion}
\label{sec:discussion} 

Our modelling of the long-term scintillation of PSR J1141$-$6545 shows that the dominant scattering region is centred at a fractional distance of $s=0.276\pm0.008$ from the pulsar, and that the scattering is slightly anisotropic with an axial ratio of $A_r=2.14\pm0.11$. We were able to provide an independent estimate of the orbital inclination angle $i=61.3\pm2^\circ$ from the anisotropic thin screen model using uniform priors on $\cos{i}$, however this was inconsistent with the value inferred from pulsar timing. We repeated the fit with a Gaussian prior on the inclination angle, to obtain improved measurements of the other correlated scintillation parameters. The measured parameters from both of these models, the derived anisotropy axial ratio $A_r$, the mass of pulsar $M_{\rm p}$, and the mass of the companion $M_{\rm c}$ are given in Table \ref{tab:bestparams}. We also derive a new estimate of the pulsar distance $D$ and proper motion in celestial coordinates for the first time, as described below in Section \ref{sec:distance}. We then discuss the implications of these derived measurements for future tests of GR in Section \ref{sec:timing}.

\subsection{Pulsar distance and proper motion estimates}
\label{sec:distance}
The distance to PSR J1141$-$6545 is currently poorly constrained, with the best estimate of $D=3\pm2$\,kpc from \citet{Verbiest+12} originating from a luminosity bias correction to an earlier lower-limit of $3.7\pm1.7$\,kpc that was derived from neutral hydrogen absorption spectrum \citep{Ord+02b}. There is also a lower estimate of 1.7\,kpc from the most recent Galactic electron-density model \citep{Yao+17}. This poor measurement precision is problematic for our data because our derivation of $V_{\rm ISS}$ from the scintillation parameters is proportional to $\sqrt{D}$. We accordingly fitted for a scaling factor $\kappa$ (see Section \ref{sec:model}), which would absorb any errors in the calculation of $V_{\rm ISS}$, for example from an error in the assumed pulsar distance of $D=3$\,kpc. However, this scaling factor would also include systematic errors in the measurements of $\Delta\nu_{\rm d}$ and $\tau_{\rm d}$, as well as any error in the numerical relationship between $\Delta\nu_{\rm d}$, the strength of scattering, and $s_{\rm d}$ derived by \citet{Cordes+98}. This numerical relationship is summarised by the $A_{\rm ISS}$ coefficient, $A_{\rm ISS}=2.78\times10^{4}\sqrt{2(1-s)/s}$\,km\,s$^{-1}$ for the thin screen model.

We have also chosen to scale each epoch of observations to the mean $V_{\rm ISS}$, which will primarily take into account any epoch-to-epoch errors in the relationship between the strength of scattering and $\Delta\nu_{\rm d}$. In Section \ref{sec:inhomogeneity} we showed that it was necessary to re-estimate $\Delta\nu_{\rm d}$ at each epoch because of inhomogeneities in the IISM on an AU scale. However we also know that the derivation of $s_{\rm d}$ from $\Delta\nu_{\rm d}$ is imperfect and can depend on the shape of the ACF, among other factors. This was also noted by \citet{Rickett+14}, who found discrepancies between their $\Delta\nu_{\rm d}$ measurements and the spatial scale inferred from measurements of the harmonic coefficient $K_{C2}$ with time. For our analysis, we intend for our scaling of each epoch to take into account such discrepancies with time, which may also come partially from time-evolution in other parameters, such as the anisotropy. However, we assume that the scaling factor for the mean $V_{\rm ISS}$, $\kappa$, represents the largest systematic error in the calculation of $V_{\rm ISS}$, and we assume that this is from an error in the pulsar distance $D$.

For the two best models, we show this scaling factor $\kappa$ and its error in Table \ref{tab:bestparams}. For our preferred model with a Gaussian inclination angle prior, $\kappa=0.496\pm 0.014$, suggesting a significant systematic error in the $V_{\rm ISS}$. If this is indeed a measurement of the error in pulsar distance, we can use this to provide a new estimate. To do so, we first assume a 20\% uncertainty on $A_{\rm ISS}$, originating from a discrepancy of approximately this magnitude between the spatial scale $s_{\rm d}$ derived from the measured $\Delta\nu_{\rm d}$, and that derived from $K_{C2}$ in this work and that of \citet{Rickett+14}. We then calculate the distribution for pulsar distance given that $D\propto1/\kappa^2$ and assume a Gaussian distribution for $\kappa$ (which is reasonable, from Figure \ref{fig:posterior_i73}) with mean and standard deviation given by the measurement in Table \ref{tab:bestparams}. We do this for both models (for comparison), and find that the distance is estimated to be larger than the assumed $3$\,kpc, but that it is poorly constrained by this method, $D=10^{+4}_{-3}$\,kpc. 

If the pulsar is this distant, it may be surprising that we observe significant anisotropy and localised scattering. But while most pulsars that have been studied with scintillation are relatively nearby, there are examples of distant thin screens observed out to of order few kpc \citep[e.g.][]{Putney+06, Popov+16}. Indeed, in this work we have precisely measured the fractional distance to a scattering screen, which would be placed at $\sim 2$\,kpc even for the lower suggested pulsar distance \citep{Verbiest+12, Yao+17}. We find no clearly associated HII regions for this line-of-sight in a search of the WISE catalogue \citep{Anderson+14} (the nearest likely region has a radius 8.6\,arcmin centred $\sim$30\,arcmin from the line-of-sight). Our estimate shows that in principle scintillation can be used to estimate pulsar distances, but in practice it is complicated by the $D\propto1/\kappa^2$ relationship and by the large uncertainty for $A_{\rm ISS}$. With improved scintillation modelling and understanding of the IISM along the line-of-sight, we expect that the precision on distance estimates with this method can be improved in the future.

Using our own distance estimate (for self-consistency) with our measurements of the pulsar's transverse velocity $v_{\mu, \parallel}$ and $v_{\mu, \perp}$, and the longitude of the ascending node $\Omega$, we have been able to derive the pulsar proper motion in celestial coordinates (right ascension $\alpha$ and declination $\delta$). We used the same distance distribution derived from $\kappa$ above, and calculated the distribution of proper motions assuming Gaussian distributions for $v_{\mu, \parallel}$, $v_{\mu, \perp}$, and $\Omega$ given by the measurements and their uncertainties listed in Table \ref{tab:bestparams}. The derived proper motions in $\alpha$ and $\delta$ are shown in the table for both models, and are consistent because of the large uncertainty. For the Gaussian $i$ prior model, we have $\mu_{\alpha}\cos{\delta}=2.9\pm1.0$\,mas\,yr$^{-1}$ and $\mu_{\delta}=1.8\pm0.6$\,mas\,yr$^{-1}$. We believe that this is a fairly conservative estimate (because of the large uncertainty on $D$), but may be under-estimated if other existing distance estimates are correct. For example, taking the current lowest distance estimate of $D=1.7$\,kpc from \citet{Yao+17} and assuming a 20\% uncertainty, we would have $\mu_{\alpha}\cos{\delta}=15\pm4$\,mas\,yr$^{-1}$ and  $\mu_{\delta}=9\pm3$\,mas\,yr$^{-1}$. However a proper motion of this magnitude may soon be ruled out with improved pulsar timing sensitivity.

The estimated proper motion is highly uncertain, primarily because of uncertainty in the pulsar distance. For the following section we will assume the values derived from our new distance estimate. However, if in the future the pulsar distance is constrained with higher confidence, the proper motion should be re-derived from our measurements of $v_{\mu, \parallel}$, $v_{\mu, \perp}$, and $\Omega$ in Table \ref{tab:bestparams}. Alternatively, given a measurement of the pulsar proper motion in the near future from improved timing precision, our independent measurements of the velocity can be used to derived the distance.

\subsection{Implications for timing and tests of gravitational theories}
\label{sec:timing}
PSR J1141$-$6545 is a highly relativistic pulsar in an eccentric, asymmetrical mass system, which makes it an ideal laboratory for testing GR. \citet{Bhat+08} analysed the gravitational radiation losses from this system through pulsar timing and noted that with increasing precision of the orbital period-derivative, contamination from kinematic effects (e.g. the Shklovskii effect) and Galactic acceleration would start to dominate the uncertainty in the near future. The transverse velocity of the pulsar system is accompanied by a radial acceleration, which produces a time-dependent Doppler-shift to the pulsar spin frequency and orbital period. This is the Shklovskii effect \citep{Shklovskii70}, and it results in an apparent orbital period-derivative $\dot{P}_b^{\rm kin}$ that is considered a contamination to the orbital period-derivative measurement from gravitational radiation losses, $\dot{P}_b^{\rm GR}$. However, the exact level of this effect was unknown because the proper motion was not measured via pulsar timing. We are now able to determine the contribution from the Shklovskii effect for the first time because our scintillation work is sensitive to the transverse motion of the pulsar instead of the radial motion probed by pulsar timing.

This Shklovskii effect can be calculated from $\dot{P}_b^{\rm kin}=DP_b\mu ^2/c = P_b V^2_\mu/cD$ \citep{Bell+96}. With the distance and transverse velocity provided by our model, we calculate $P_b^{\rm kin}=3.5\times10^{-15}$, which is $\sim$1\% of the $\dot{P}_b$ measurement of \citet{Bhat+08} and well below the 6\% measurement precision. However, if we again assume the distance from \citet{Yao+17}, then $\dot{P}_b^{\rm kin}=1.9\times10^{-14}$. This is at the level of the current expected timing precision of $\sim$2\% for $\dot{P}_b$. It is therefore important to further constrain the system transverse velocity and/or distance from improved scintillation modelling and/or pulsar timing. In addition to the Shklovskii effect on $\dot{P}_b$, the proper motion changes the projected geometry of the binary orbit, resulting in apparent $\dot{x}$ and $\dot{\omega}$. However, these kinematic contaminations are far below the measurement precision for this pulsar and are typically only observed in precisely-timed millisecond pulsars \citep{Arzoumanian+96,Kopeikin96}. 

The ongoing long-term timing campaign on this system will also benefit from the robust measurements of the sense of the inclination angle and $\Omega$ obtained from this work. These measurements will help understand relative contributions to the secular variations of the orbit from phenomena such as relativistic aberration, spin-orbit coupling, and proper motion. Measurements of $\Omega$ will also be helpful in understanding the pulsar emission beam geometry with the precessional evolution of its polarisation \citep{Kramer+09}. Measurements of $\Omega$ combined with $\dot{\omega}$ from timing, can be used to test preferred-frame effects predicted by a variety of alternative theories of gravity \citep{Wex+07, Shao+12}.

\section{Conclusion}
We have presented new scintillation models for PSR J1141$-$6545 using six years of data from the Parkes 64\,m radio telescope. We found that like many pulsars, the scattering shows some anisotropy, and is dominated by a single scattering region centred at $s=0.276\pm0.008$. By accounting for anisotropy in the scattering, we measured the system inclination angle of $i=61.3\pm 2^\circ$, which is significantly lower than the constraint of $73\pm3^\circ$ from pulsar timing (inferred from the pulsar and companion masses derived with GR), suggesting that a weakness in the model, such as time-stationary IISM velocity and anisotropy may cause some systematic errors. However, by using this pulsar timing constraint as a prior probability distribution, we have been able to measure several astrometric parameters for the first time.

Using the significant annual and relativistic variations observed in the scintillation velocity to constrain long-term models, we have been able to resolve the ``sense" of the inclination angle, and we find that $i<90^\circ$. This in turn resolved the ambiguity in the direction of the proper motion velocity in pulsar coordinates. With our new measurement of the orientation of the orbit in celestial coordinates $\Omega=24.8\pm1.8^\circ$, and estimate of the pulsar distance $D=10^{+4}_{-3}$\,kpc, we have been able to estimate the proper motion for the first time. We determine the proper motion in right ascension $\mu_{\alpha}\cos{\delta}=2.9\pm1.0$\,mas\,yr$^{-1}$ and in declination $\mu_{\delta}=1.8\pm0.6$\,mas\,yr$^{-1}$, and we use these numbers to calculate the contribution of the Shklovskii effect to $\dot{P}_b$. This effect is the most significant source of contamination for tests of GR, but our low proper motion suggests that it exists only at the $\sim$1\% level. Our improved accuracy and precision for the pulsar's transverse velocity is also important for understanding the formation of this system.

Our distance measurement is model-dependent, with the accuracy determined partially by the relationship between the measured scintillation bandwidth and the spatial scale, for which we have used the method of \citet{Cordes+98}. In the near future, wide-bandwidth observing systems, such as that of the MeerKAT radio telescope or the ultra-wideband low-frequency (UWL) receiver for the Parkes 64\,m radio telescope, could provide an experimentally-derived relationship through analysis of the frequency-dependence on the scintillation bandwidth. In this way scintillation studies may be used to give distance measurements to pulsars with predictable modulation of the scintillation timescale (e.g. relativistic binaries). 

We may also soon see an independent proper motion measurement from pulsar timing with improved observation span and techniques, which would provide another method for determining the pulsar distance in combination with our velocity measurements. Future high-quality observations, with the sensitivity and resolution to estimate the scattering anisotropy directly from the autocovariance functions of individual dynamic spectra \citep[as in][]{Rickett+14} will be valuable for improving the accuracy and precision of future inclination angle measurements of scintillating binary pulsars.

\section*{Acknowledgements}
We thank Barney Rickett for useful discussion and for allowing us access to code and data used for the earlier work in \citet{Rickett+14}. The Parkes radio telescope is part of the Australia Telescope National Facility which is funded by the Commonwealth of Australia for operation as a National Facility managed by CSIRO. Work at NRL is supported by NASA. MB is supported by the ARC Centre of Excellence OzGrav under grant CE170100004.



\bibliographystyle{mnras}
\input{Complete_Manuscript_File.bbl}




\section*{Appendix A: Scintillation velocity models}
\label{sec:AppendixA}

From the dynamic spectrum of intensity scintillations as a function of frequency and time, one can derive a scintillation velocity $V_{\rm ISS}$ (Equation \ref{eqn:viss}) that relates to the structure and velocity of the diffraction pattern drifting across the observer. A line-of-sight velocity model for $V_{\rm ISS}$ is found by integrating $V_{\rm eff}$ (Equation \ref{eqn:veff1}) along the line-of-sight from $x=0$ at the pulsar and $x=1$ at the observer, with a weight that corresponds to the geometry of the scattering medium \citep{Cordes+98}

\begin{equation}
\label{eqn:viss_int}
V_{\rm ISS} = \left[\frac{\int\limits_0^1 \eta(x)\lvert V_{\rm eff}(x)\rvert^\alpha dx}{\int\limits_0^1 \eta(x)x^\alpha dx}\right]^{1/\alpha},
\end{equation}

\noindent where $\alpha=5/3$ for a Kolmogorov medium, and $\eta(x)$ is the mean-square scattering angle per unit distance and functions as the weight for the integral to describe different geometries. For uniform scattering along the line-of-sight $\eta(x)=1$, while for a thin screen $\eta(x)$ is a delta function at $x=s$ and the expectation of $V_{\rm ISS}$ reduces to the line-of-sight velocity with respect to the diffraction pattern $V_{\rm los}$, defined at the location of the observer,

\begin{equation}
\label{eqn:viss_screen}
E(V_{\rm ISS}) = V_{\rm los} = V_{\rm eff}/s, 
\end{equation} 

\noindent where $E(V_{\rm ISS})$ denotes the expectation of the observationally-derived $V_{\rm ISS}$, and $V_{\rm{eff}}$ is given in Equation \ref{eqn:veff1}. The transverse orbital velocity of the pulsar $V_{\rm p}$ has components $v_{{\rm p},\parallel}$ along the line of nodes, and $v_{{\rm p},\perp}$ perpendicular to this in the plane of the sky. These velocities are defined as a function of orbital phase from the line of nodes $\phi=\theta+\omega$, where $\theta$ is the true anomaly and $\omega$ is the longitude of periastron,
\begin{equation}
\label{eqn:vcomponents}
\begin{aligned}
v_{{\rm p},\parallel} &= -V_0\left(e\sin{\omega} + \sin{\phi}\right) \\
v_{{\rm p},\perp} &= V_0\cos{i}\left(e\cos{\omega} + \cos{\phi}\right),
\end{aligned}
\end{equation}
\noindent where $V_0=2\pi xc/(\sin{i}P_b (1 - e^2)^{1/2})$ is the mean orbital velocity, $x$ is the projected semi-major axis in seconds, $P_b$ is the binary orbital period, $e$ is the eccentricity, $i$ is the inclination angle. The true anomaly is first calculated by numerically computing the eccentric anomaly, $E$ from Kepler's equation $E-e\sin{E}=M$, with mean anomaly $M=(2\pi/P_b)(t-T_0)$, where $T_0$ is the epoch of periastron. The true anomaly $\theta$ is then given by 
\begin{equation}
\label{eqn:true anomaly}
\theta = 2\arctan\left[\sqrt{\frac{1+e}{1-e}}\tan{\frac{E}{2}}\right].
\end{equation}
The known contribution of the Earth's velocity is then rotated from celestial coordinates to the pulsar-frame coordinates according to the longitude of the ascending node $\Omega$ for the pulsar's orbit
\begin{equation}
\begin{pmatrix} v_{\rm E, \parallel} \\ v_{\rm E, \perp}  \end{pmatrix} = \begin{pmatrix} \sin{\Omega} & \cos{\Omega} \\ \cos{\Omega} &-\sin{\Omega}  \end{pmatrix}\begin{pmatrix} v_{{\rm E}, \alpha} \\ v_{{\rm E}, \delta}  \end{pmatrix}
\end{equation}

\noindent where $v_{\rm E, \parallel}$ and $v_{\rm E, \perp}$ are the components of the Earth's velocity aligned with $v_{{\rm p},\parallel}$ and $v_{{\rm p},\perp}$ respectively, and $\Omega$ is defined as a rotation East of North. This was also used by \citet{Rickett+14} (but with the direction of the perpendicular axis reversed in their definition) to include the Earth's velocity and proper motion of PSR J0737$-$3039A into their scintillation model. We then combine the pulsar and Earth velocities, include the pulsar system transverse velocity ($v_{{\mu},\parallel}$ and $v_{{\mu},\perp}$), and scale them appropriately by the distance to the scattering region $s$

\begin{equation}
\begin{aligned}
v_{\parallel}(s) &=  s v_{{\rm E},\parallel} + (1-s) (v_{{\rm p},\parallel} + v_{{\mu},\parallel}) \\ v_{\perp}(s) &=  s v_{{\rm E},\perp} + (1-s)( v_{{\rm p},\perp} +v_{{\mu},\perp}) .
\end{aligned}
\end{equation}

In the case of isotropic scattering where the angular size of the pulsar orbit is compact enough to remain in the scattering disk, the spatial scale $s_{\rm d}$ is constant with orbital phase (but it may change on longer time scales if the IISM is inhomogeneous; see Section \ref{sec:inhomogeneity}). Consequently, the decorrelation bandwidth $\Delta\nu_{\rm d}$ is also constant with orbital phase and the effective velocity is simply given by 

\begin{equation}
\label{eqn:veff_isotropic}
V_{\rm eff}(s)=\sqrt{v_{\parallel}(s)^2 + v_{\perp}(s)^2}.
\end{equation}

However, for anisotropic scattering, $s_{\rm d}$ depends on the direction of $V_{\rm ISS}$. To account for such scattering in PSR J0737$-$3039A, \citet{Coles+05} considered the spatial diffraction pattern as an ellipse. The pattern is then described by a quadratic form $Q(\mathbf{s_{\rm d}})=as_{\rm d,\parallel}^2 + bs_{\rm d,\perp}^2 + cs_{\rm d,\parallel}s_{\rm d,\perp}$, where the coefficients $a$, $b$, and $c$ are parametrised by the axial ratio $A_r$ of the ellipse and its orientation $\psi$ with respect to the coordinates of the pulsar orbit as defined above. \citet{Rickett+14} used this anisotropy model for PSR J0737$-$3039A but parametrised the quadratic coefficients in terms of $R=(A_r^2 - 1)/(A_r^2 + 1)$, which is bound between 0 and 1. If the orientation angle $\psi$ is defined clockwise from the line of nodes, then from \citet{Rickett+14} the coefficients are

\begin{equation}
\label{eqn:anisotropy}
\begin{aligned}
a &=  \left[1 - R\cos{\left(2\psi\right)}\right]/\sqrt{1 - R^2} \\ 
b &=  \left[1 + R\cos{\left(2\psi\right)}\right]/\sqrt{1 - R^2} \\ 
c &=  - 2R\sin{\left(2\psi\right)}/\sqrt{1 - R^2}.
\end{aligned}
\end{equation}

Finally, we introduce a scaling factor $\kappa$ to the model, which will account for any errors (for example an error in the pulsar distance $D$) in the calculation of $V_{\rm ISS}$ from the dynamic spectrum (Equation \ref{eqn:viss}). Our final model for the effective velocity is then

\begin{equation}
\label{eqn:veff_anisotropic}
V_{\rm eff}(s) = \kappa\sqrt{a v_{\parallel}(s)^2 + b v_{\perp}(s)^2 + c v_{\parallel}(s)v_{\perp}(s)},
\end{equation}

\noindent which we use in Equation \ref{eqn:viss_int} with $\eta(x)=1$ for the uniform medium model and Equation \ref{eqn:viss_screen} for the thin screen model.

This anisotropic scattering model can alternatively be represented as a sum of five harmonics, which we used to model each of the individual epochs. This model is described in Section \ref{sec:shortterm}, but the full dependence on physical parameters for each of the harmonic coefficients in Equation \ref{eqn:harmcoeffs} is given below \citep{Rickett+14}

\begin{align}
\label{eqn:normharms}
K_0=& [0.5V_0^2(a+b\cos^2{i})\\ \nonumber
& + a(v_{\rm C,\parallel} - V_0e\sin\omega)^2 \\ \nonumber
& +b(v_{\rm C,\perp} +V_0e\cos\omega\cos{i})^2\\ \nonumber
& +c(v_{\rm C,\parallel}-V_0e\sin\omega)(v_{\rm C,\perp}+V_0e\cos\omega\cos{i})]/s_{\rm d}^2\\
  \nonumber
K_S=& -V_0[2a(V_{\rm C,\parallel} - V_0e\sin\omega) \\ \nonumber
& + c(v_{\rm C,\perp}+V_0 e\cos{i}\cos\omega)]/s_{\rm d}^2 \\ \nonumber
K_C=&V_0\cos{i}[c(v_{\rm C,\parallel}-V_0e\sin\omega) \\ \nonumber
& +2b(v_{\rm C,\perp}+V_0e\cos{i}\cos\omega)]/s_{\rm d}^2 \\ \nonumber
K_{S2}=&-cV_0^2\cos{i}/2s_{\rm d}^2 \\ \nonumber
K_{C2}=& V_0^2(-1+\cos^2{i})/2s_{\rm d}^2\nonumber
\end{align}
\noindent where $v_{\rm C,\parallel}$ and $v_{\rm C,\perp}$ are the constant (for a given epoch) components of the line-of-sight velocity, which will be dominated by the pulsar's proper motion, but also includes the Earth's velocity and any IISM velocity.

\section*{Appendix B: Reproducing our results}
\label{sec:AppendixB}

The raw data from the Parkes radio telescope that were used for this work are available from the CSIRO data access portal (DAP;
\url{https://data.csiro.au}), and were processed with a pipeline developed for the second PPTA data release. This pipeline is used to produce TOAs and dynamic spectra, and is briefly summarised in Section \ref{sec:dataset}. The dynamic spectra files that were produced from this pipeline, and the collection of MATLAB codes used to analyse these dynamic spectra for the results presented in this paper, are available from the DAP at \url{https://doi.org/10.4225/08/5ae7b7b1b65c8}. 

The code is presented for reproducibility and in general is not intended to be used for other applications, however a brief description of each script and function is included in \textsc{README.txt}, and some may be useful for other scintillation studies. For example, we have included a function (\textsc{getDynspecParams.m}) that may be used to measure scintillation parameters for any dynamic spectrum given as a two-dimensional matrix of intensity versus observing time and frequency, using the methods described in Section \ref{sec:measureparams}. We have also included the scintillation velocity model (\textsc{vissmodel.m}), which is described in Section \ref{sec:model}, and a script that numerically calculates the true anomaly from MJD and then uses nonlinear regression to fit the model to $V_{\rm ISS}$ derived from scintillation parameters (\textsc{modelDynspecParams.m}). 

The MATLAB code was executed in version 2017b, and requires \textsc{jd2date.m} from the \textsc{Astromatlab} library \citep[\url{https://webhome.weizmann.ac.il/home/eofek/matlab/};][]{Ofek14}, MATLAB Multinest \citep{Pitkin+12}, and the ``Image Processing", ``Statistics and Machine Learning", ``Signal Processing", ``Optimization", ``Curve Fitting", ``Symbolic Math", and (optionally) ``Parallel Computing" MATLAB toolboxes. 


\bsp	
\label{lastpage}
\end{document}